\pdfoutput=1
\documentclass[journal,12pt,onecolumn]{IEEEtran}

\usepackage{amsmath,graphicx,epsfig,float,epstopdf,subfigure,times,latexsym,verbatim,mathrsfs,amssymb,textcomp,txfonts,color,mathrsfs}
\usepackage[flushleft]{threeparttable}
\usepackage{algorithm}
\usepackage{algorithmic}
\usepackage{url}

\renewcommand{\baselinestretch}{0.97}
\IEEEoverridecommandlockouts

\begin{document}
\title{{M$^2$I: Channel Modeling for Metamaterial-Enhanced\\ Magnetic Induction Communications}}

\author{Hongzhi Guo, Zhi Sun, Jingbo Sun, and Natalia M. Litchinitser% <-this % stops a space
\thanks{The authors are with the Department of Electrical Engineering, University at Buffalo, the State University of New York, Buffalo, NY 14260, USA.}
\thanks{Zhi Sun handles the correspondence of this paper. Phone number: +1 (716) 645-1608. Fax number: +1 (716) 645-3656. E-mail: zhisun@buffalo.edu.}
\thanks{This work was supported by the US National Science Foundation (NSF) under Grant No. 1547908.}% <-this % stops a space
\vspace{-20pt}
}

%\markboth{IEEE Transactions on Wireless Communications}%
%{Submitted paper}

%\IEEEpubid{0000--0000/00\$00.00\copyright~2007 IEEE}

\maketitle
\begin{abstract}

Magnetic Induction (MI) communication technique has shown great potentials in complex and RF-challenging environments, such as underground and underwater, due to its advantage over EM wave-based techniques in penetrating lossy medium.
However, the transmission distance of MI techniques is limited since magnetic field attenuates very fast in the near field.
To this end, this paper proposes Metamaterial-enhanced Magnetic Induction (M$^2$I) communication mechanism, where a MI coil antenna is enclosed by a metamaterial shell that can enhance the magnetic fields around the MI transceivers.
As a result, the M$^2$I communication system can achieve tens of meters communication range by using pocket-sized antennas.
In this paper, an analytical channel model is developed to explore the fundamentals of the M$^2$I mechanism, in the aspects of communication range and channel capacity, and the susceptibility to various hostile and complex environments. The theoretical model is validated through the finite element simulation software, Comsol Multiphysics. Proof-of-concept experiments are also conducted to validate the feasibility of M$^2$I.
%The developed channel model and analysis provide principles and guidelines in designing the M$^2$I communication systems.

%The magnetic field intensity generated by the coil are developed and validated by software simulation. Then the channel characteristics of the point-to-point communication and Magnetic induction waveguide are found. Moreover, the practical metamaterial loss, dispersion, and coil's orientation are discussed.
\end{abstract}

\IEEEpeerreviewmaketitle

\vspace{-10pt}

\section{Introduction}
Despite the presence of wireless connectivity in most terrestrial scenarios, there are still many hostile and complex environments that cannot be covered by existing wireless communication techniques, including underground, underwater, oil reservoirs, groundwater aquifers, nuclear plants, pipelines, tunnels, and concrete buildings.
Wireless networks in such environments can enable important applications in environmental, industrial, homeland security, and military fields, such as monitoring and maintenance of groundwater and/or oil reservoirs \cite{Guo_oil_2014}, or damage assessment and mitigation in nuclear plants \cite{nawaz2010underwater}, among others.
However, the harsh wireless channels prevent the direct usage of conventional electromagnetic (EM) wave-based techniques due to the high material absorption when penetrating lossy media. %\textcolor{red}{here we may add "due to the high material absorption (lossy medium), reflection, and multiple scattering (complicated structure)", magnetic field also experiences absorbtion.}

Among potential solutions, the Magnetic Induction (MI) technique has shown great potentials in underground \cite{Akyildiz_WUSN_Survey_2006} and underwater \cite{Gulbahar_underwater} environments. In a MI communication system, the HF band magnetic field generated by a MI transmitter coil is utilized as the signal carrier \cite{Sun_MI_TAP_2010}. Since most natural media have the same magnetic permeability as air, MI keeps the same performance in most materials. Even in lossy media like groundwater, the MI path loss caused by skin depth can be minimized since MI communication is realized within one wavelength from the transmitter \cite{Sun_TCOM_2013}. In addition, MI does not suffer from the multipath fading problem in EM wave-based solutions \cite{Gulbahar_underwater}. However, MI systems depend on the magnetic field generated by the transceivers in the near field, which attenuates very fast. Consequently, the range of MI communication is very limited. %unless very large coil antenna and high transmission power are used \cite{Gulbahar_underwater}, which are infeasible in the aforementioned applications.

To this end, we introduce metamaterials to MI communications, which can manipulate and enhance the magnetic fields transmitted and received by MI transceivers. %\textcolor{red}{here we already defined the concept of metamaterial}
Metamaterials are artificial structures made of carefully designed building blocks, which can generate unique physical phenomenon such as backward waves and negative refraction index \cite{Veselago_Meta_1968, Pendry_2000}. The novel properties of metamaterials have been utilized in subwavelength imaging \cite{Freire_MRI}, wireless power transfer \cite{Wang_WPT}, and antenna miniaturization \cite{Lopez_LTE}.
Since the key problem of the MI communication technique is the fast fall-off rate in near field, we see great potentials in using metamaterials to enhance MI-based communications and finally achieve both extended medium penetration performance and practical communication ranges.
To date, no efforts have explored the design of the metamaterial enhancement of MI communications in complex environments.

\begin{figure}%[H]
  \centering
    \includegraphics[width=0.4\textwidth]{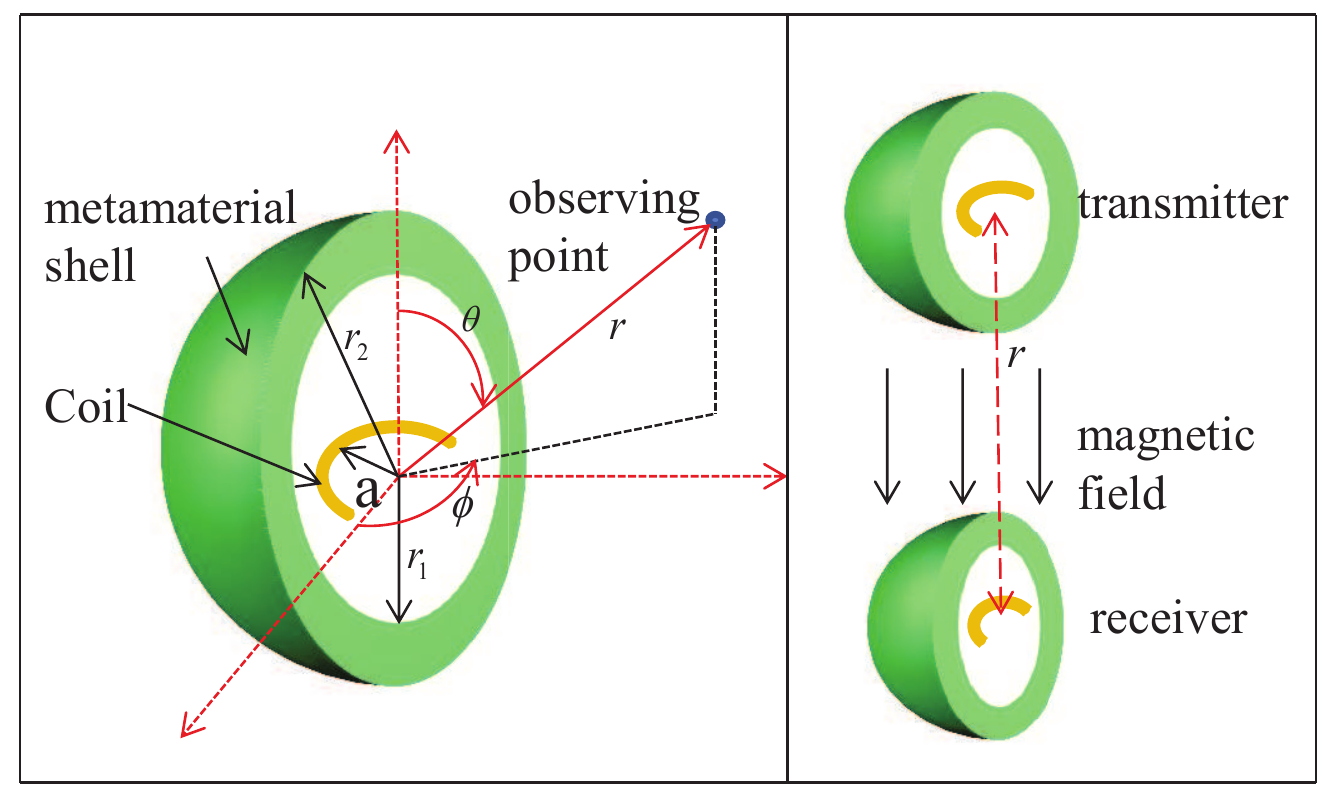}
    \vspace{-5pt}
  \caption{Illustration of the M$^2$I communication between two M$^2$I transceivers (MI coil enclosed in a metamaterial spherical shell).}
  \vspace{-15pt}
  \label{fig:sys}
\end{figure}

In this paper, we propose the Metamaterial-enhanced Magnetic Induction (M$^2$I) communication mechanism for the aforementioned wireless applications in various environments that are structurally complex and challenging for RF wireless signals.
By introducing the M$^2$I antennas (the MI coil antenna enclosed by a metamaterial-enabled resonant sphere, as shown in Fig.~\ref{fig:sys}), we show that the efficient wireless communication can be realized in lossy environments with good range. The whole communication process (starting from the transmitter, via the lossy transmission medium, and ending at the receiver) is investigated as an integrated system.
We develop an analytical channel model that quantitatively captures the unique interactions among MI transceivers, the metamaterial-enabled resonant structure, and complex environments, which are not observed in existing metamaterial applications.
The proposed M$^2$I mechanism and the channel model are validated by both the Finite Element Method (FEM) software, i.e., Comsol Multiphysics \cite{comsol}, and proof-of-concept experiments. Based on the derived channel model, we confirm the feasibility of achieving tens of meters communication range in M$^2$I systems by using pocket-sized antennas.
The developed channel model bridges the communication system optimization with the metamaterial device design.

%In particular, an analytical channel model is developed to explore the fundamentals of the M$^2$I technique under the impacts from environments. Explicit and tractable expressions of the path loss and bandwidth of the M$^2$I channel are developed by analyzing: (i) the novel properties of the complex self-inductance and mutual inductance introduced by M$^2$I, (ii) the metamaterial-manipulated EM field around the M$^2$I transceivers, and (iii) the joint effects of the metamaterial resonance and the spatial resonance in M$^2$I. The M$^2$I performances, including the communication range, the channel capacity, the system robustness and reliability, and the susceptibility to various hostile and complex environments, are investigated using the developed channel model.

The reminder of this paper is organized as follows. Related works are presented in Section II. Then, in Section III, an analytical channel model for M$^2$I communication is developed to characterize how the metamaterial sphere works in M$^2$I systems in lossy media. Next, the channel characteristics of M$^2$I communication including the point-to-point and MI waveguide communication are discussed in Section IV. Finally, this paper is concluded in Section V.

\vspace{-5pt}
\section{Related  Work}
%\vspace{-5pt}

MI techniques have been utilized in many complex environments.
%mines/tunnels
In \cite{hjelmstad1991ultra, Reagor_super, vasquez2004underground}, voice and low data rate communications have been established by MI in underground mines.
%underwater
In \cite{Gulbahar_underwater,Sojdehei_MI_Com_2000, wrathall2000magneto}, MI communication is realized in lossy underwater environment, where very large coil antennas are utilized.
%underground
In \cite{Akyildiz_WUSN_Survey_2006, Sun_MI_TAP_2010, Parameswaran_MI}, MI is introduced to wireless underground sensor networks, where wirelessly networked sensor devices are buried in soil medium.
%intrabody
In \cite{Puers_Capsule}, MI is utilized to transmit both data and power into human body for medical applications.
%commercial
Besides theoretical research, many commercial MI systems have also been developed for mining safety and undersea surveillance \cite{vitalalert, Lockheed_MI_2012, ultra-ms}, among others.
Despite of the advantages, the existing MI communication systems have very limited ranges due to the fast fall-off in near field unless very large coil antennas are used.
To extend the very limited range, waveguide structures \cite{Shamonina_Waveguide} can be utilized. In \cite{Sun_MI_TAP_2010}, we show underground MI communication range can be significantly extended by passive relay coils, i.e., the MI waveguides. However, existing MI waveguides require very high density of relay coils, which prevents practical implementation.

%It is possible to utilize such novel property to address the aforementioned problems in MI and MI waveguide techniques.

Metamaterials have been utilized in a wide range of applications, such as the metamaterial cloak, metamaterial enhanced MRI \cite{Freire_MRI}, and metamaterial antenna \cite{Lopez_LTE}.
Among the various research thrusts in metamaterials, two areas are most relevant to our work, including wireless energy transfer and RF antenna miniaturization.
(i) In \cite{Wang_WPT}, a metamaterial slab is introduced to increase the wireless energy transfer efficiency. In \cite{Urzhumov_Meta, Urzhumov_experiment}, the metamaterial enhancement in energy transfer is validated in both theoretical analysis and experiments. Different from the single frequency wireless energy transfer, the M$^2$I communication system proposed in this paper requires the wireless signals occupy a significant bandwidth to carry information with high date rate. Moreover, existing metamaterial-enhanced wireless energy transfer systems need a large metamaterial slab (much larger than the coil itself). The charging range is too short for communication systems. Therefore, a technological breakthrough is required to realize the M$^2$I communication.
(ii) In \cite{Engheta_cavity}, metamaterials are introduced to the field of RF antenna miniaturization. In \cite{Ziolkowski_electric, Arslanagic_Sphere}, an electrical dipole antenna enclosed in a metamaterial shell is investigated in deep subwavelength range. The far field propagating wave and the radiated power from the electrical dipole can be dramatically amplified in lossless air medium. In contrast, the M$^2$I communication discussed in this paper focuses on the near field EM components, especially the magnetic field around magnetic dipole (i.e., coil), which needs a major reexamination on the metamaterial resonant structure. More importantly, M$^2$I communication is designed to operate in complex media, which dramatically change the condition and the properties of metamaterial resonance. In addition, when comparing performances with conventional antennas, existing works, such as \cite{Ziolkowski_electric}, use the same dipole moments. However, since M$^2$I in lossy medium can have very large frequency-dependent resistance, the metamaterial antenna used in M$^2$I can have dramatically different dipole moments from conventional antennas when the input power are the same.

\section{Modeling and Analysis of M$^2$I Communications}
\label{sec:channel_model}
In this section, an analytical channel model of the proposed M$^2$I communication technique is developed for complex and RF-challenging environments. Specifically, metamaterials are introduced to enhance both the wireless communications using point-to-point MI and MI waveguide. The MI waveguide is actually a sequence of point-to-point pairs. Hence it shares the same theoretical foundation as point-to-point MI. Therefore, we first developed the path loss model for point-to-point M$^2$I, which can be easily extended to M$^2$I waveguide. The discussion on the optimal metamaterial shell configuration is also universal for both settings.

In the following analysis, we use boldface lowercase letters for vectors and boldface capital letters for matrices. For a vector ${\bf h_{\theta}}$, we use $h_{\theta}$ to denote its magnitude and a unit vector $\hat{\theta}$ to denote its direction. For a matrix ${\bf S}$, ${\bf S}^{t}$ denotes its transpose and $\det({\bf S})$ denotes its determinant. For a complex number, we use $\Re(\cdot)$ and $\Im(\cdot)$ to denote the real and the imaginary parts, respectively. If there is no special notation, the considered lossy medium in this paper is underground soil medium.

%\textcolor{red}{First, the mechanism of the M$^2$I communication is discussed based on the channel models of the original MI communication (including MI waveguide) techniques \cite{Sun_MI_TAP_2010}. %The novel property of the M$^2$I communication is characterized.
%Then, the electromagnetic field around the M$^2$I transmitter and receiver is analyzed. %Based on the field analysis, two important parameters in M$^2$I can be numerically derived, including the self-inductance of a single M$^2$I transceiver and the mutual inductance between two M$^2$I transceivers.
%Finally, the analytical channel model of M$^2$I Communications is developed.}%, based on which the resonance condition and the optimal configuration of $M^2I$ communication are discussed.
\vspace{-10pt}

\subsection{Enhancing MI Communication using Metamaterials}
\label{sec:circuit_model}
The original MI communication is accomplished by using a transmitter MI coil and a receiver MI coil \cite{Sun_MI_TAP_2010}. Instead of using the widely used propagating EM waves in the far field, the MI communication utilizes the magnetic field generated from the transmitter MI coil in the near field. Modulated by digital signals, such magnetic field can induce the current that carries signals at the receiver MI coil, which complete the wireless communication.

In the M$^2$I system, the MI communication is enhanced by enclosing the original MI coils with metamaterial spherical shells. The near field EM components can be manipulated and possibly enhanced by letting waves travel through the metamaterial layer. In this subsection, we initiate the analysis by discussing how the metamaterial sphere influences the original MI communication mechanism.

%\begin{figure}
%  \centering
%  \subfigure[]{
%    \label{fig:circuit}
%    \includegraphics[width=0.18\textwidth]{fig/revised_circuit}}\quad
%  \subfigure[]{%
%    \includegraphics[width=0.27\textwidth]{fig/revised_wg_circuit}
%    \label{fig:wg_circuit}}
%      \vspace{-5pt}
%  \caption{Equivalent circuit model for (a). point-to-point M$^2$I communication and (b). M$^2$I waveguide.}
%    \vspace{-5pt}
%  \label{fig:equ_circuit}
%\end{figure}

\begin{figure}%[H]
  \centering
    \includegraphics[width=0.22\textwidth]{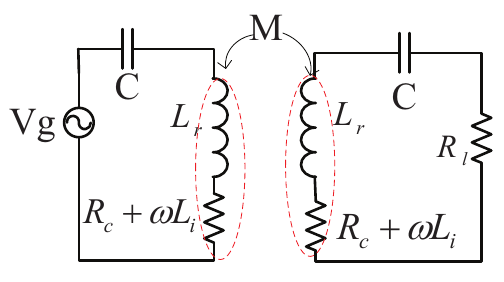}
    \vspace{-5pt}
  \caption{Equivalent circuit model for point-to-point M$^2$I communication.}
  \vspace{-10pt}
  \label{fig:circuit}
\end{figure}

According to our previous work in \cite{Sun_MI_TAP_2010}, the channel of the original MI communications (as well as the MI waveguide) can be modeled by the equivalent circuits shown in Fig.~\ref{fig:circuit}, where $R_c$ is the coil's resistance; $L_r$ is the coil's real self-inductance; $C$ is the compensation capacitor used to tune the circuit to be resonant; $R_l$ is the receiver load, $V_g$ is the source's voltage, $M$ is the mutual inductance between two adjacent coils. The ideal power source has no impedance, which is consistent with the following Comsol FEM simulations where the source is also ideal.
In order to compensate the real inductance and achieve circuit resonance, the compensation capacitor $C=\frac{1}{\omega_0^2 L_r}$ is utilized, where $\omega_0$ is the resonant frequency of the coil. It should be noted that for original MI, the imaginary part of the self-inductance $L_i\simeq$0 since there is no strong metamaterial amplification.

In M$^2$I, the equivalent circuit models are still valid but need a major modification. In particular, the metamaterial sphere can dramatically change the properties of all the self-inductances $L$ and mutual inductances $M$ in Fig.~\ref{fig:circuit}. On the one hand, the mutual inductance is expected to be dramatically increased in M$^2$I. Since the design objective of the metamaterial sphere is to amplify the near field components of EM waves, more magnetic flux goes through the transmitter and receiver coils so that the mutual inductance is increased. On the other hand, while the self-inductance $L$ is real in the original MI, it becomes a frequency-dependent complex number in M$^2$I, i.e., $L=L_r-j L_i$ where $L_r$ and $L_i$ are real positive numbers. The imaginary part of $L$ (i.e., $L_i$) forms the frequency-dependent resistance in M$^2$I antenna, which comes from two unique sources in M$^2$I: the metamaterials and the complex environments. Firstly, on the resonant metamaterial sphere that is very close to the MI coil itself, significant eddy currents are induced on the metallic components. The eddy current generates a secondary magnetic field that opposes the primary field, which reduces the current in the MI coil and equivalently increases the impedance. Secondly, the lossy medium also contribute to the imaginary part of the self-inductance $L$ due to the induced eddy current. Since the impedance of an inductor is $Z_L=j \omega L$, where $\omega$ is the angular frequency, the updated impedance with imaginary inductance is $j \omega (L_r-j L_i)=j\omega L_r +\omega L_i$. Accordingly, the compensation capacitor becomes $C=\frac{1}{\omega_0^2 L_r}$ to achieve the magnetic resonance.

Once the reactance is canceled at both transmitter and receiver, the receiver load $R_l$ is matched with the coil resistance $R_c$ and the additional loss $\omega L_i$, i.e., $R_l=R_c+\omega L_i$. Different from the EM wave-based wireless systems, the transmitter and receiver in M$^2$I and MI are closely coupled to each other so that the impudence matching are done in an integrated transmitter-receiver system. Or in another word, receiver is part of the loads in transmitter while the transmitter is the source in the receiver.

\begin{figure}%[H]
  \centering
    \includegraphics[width=0.4\textwidth]{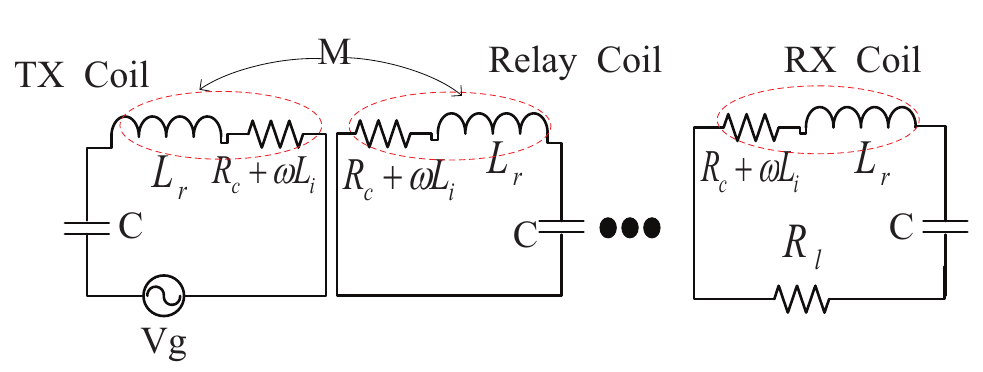}
    \vspace{-5pt}
  \caption{Equivalent circuit model for M$^2$I waveguide.}
  \vspace{-12pt}
  \label{fig:wg_circuit}
\end{figure}

Since the range of a wireless communication system is mainly determined by the channel path loss, we pay special attention to investigate the path loss in M$^2$I. It should be noted that the formulation of other important parameters in M$^2$I, including the channel bandwidth and channel capacity, can also be derived based on the path loss analysis, as shown in Section III-C and Section IV.
Based on the equivalent circuit model in Fig.~\ref{fig:circuit} and above discussions, the general path loss formula in M$^2$I channel between two transceivers (point-to-point communication) can be expressed as
\begin{subequations}
\begin{align}
\label{equ:p2p_pathloss}
\frac{P_r}{P_t}\approx\frac{\omega^2 |M|^2 R_l/({R_l+R_c+\omega L_i})}{(R_c+\omega L_i)(R_c+R_l+\omega L_i)+\omega^2 |M|^2};
\end{align}
\begin{align}
\label{equ:p2p_approx}
\mathcal{L}_{p2p}&=-10\log\left\{\frac{P_r}{P_t}\right\}\approx-20\log\frac{\omega|M|}{2(R_c+\omega L_i)},
\end{align}
\end{subequations}
where the load resistance $R_l$ is designed to maximize \eqref{equ:p2p_pathloss}, which equals $R_c+\omega_0 L_i$. To elucidate the physics better, \eqref{equ:p2p_pathloss} can be approximated by \eqref{equ:p2p_approx} at the resonant frequency. The precondition is that $\omega M$ is much smaller than $R_c+\omega L_i$. This approximation is practical since $L_i$ is very large due to the resonance of the metamaterial shell. Moreover, since we consider loose coupling for long distance communication, the mutual inductance is very weak. As a result, $\omega M$ is much smaller than $R_c$, so that the precondition holds.

%\begin{figure*}[]
%\begin{equation}
%\mathcal{L}_{p2p}=-10\log\left\{\frac{\omega^2 |M|^2 R_l/(R_l+R_c+j\omega (L_r \!-\! j L_i)+\frac{1}{j\omega C})}{(R_c+j\omega (L_r \!-\! j L_i)+\frac{1}{j\omega C})(R_c+R_l+j\omega (L_r \!-\! j L_i)+\frac{1}{j\omega C})+\omega^2 |M|^2}\right\}
%\label{equ:p2p_pathloss}
%\end{equation}
%\hrulefill
%\end{figure*}

Similarly, the M$^2$I waveguide is formed by adding the metamaterial sphere on each MI coil (including the transmitter, receiver, and relays) in an MI waveguide. Consider that there are $n-2$ relaying coils in the waveguide. The first coil is the transmitter and the last one is the receiver. All the relays are well placed along a line with equal interval $d$. Based on the equivalent circuit for M$^2$I waveguide shown in Fig. \ref{fig:wg_circuit}, the general path loss formula can be approximately expressed as
\begin{align}
\label{equ:wg_pathloss}
\mathcal{L}_{wg}\approx20(1-n)\log\left(\frac{\omega|M|}{R_c+\omega L_i}\right).
\end{align}

According to \eqref{equ:p2p_approx} and \eqref{equ:wg_pathloss}, the path loss in both M$^2$I point-to-point communication and M$^2$I waveguide are strong functions of the mutual inductance $M$, coil resistance $R_c$, and the frequency-dependent resistance $L_i$ brought by the metamaterials and the propagation medium. Since $R_c$ is a constant value, only $M$ and $L_i$ can be manipulated by designing the metamaterial sphere.
To reduce the path loss and enlarge the communication range in M$^2$I, a straightforward strategy is to increase the mutual inductance $M$ and decrease the frequency-dependent resistance $L_i$, according to \eqref{equ:p2p_approx} and \eqref{equ:wg_pathloss}. However, such strategy cannot be easily applied since the metamaterial shell can amplify both $M$ and $L_i$ simultaneously. If we reduce $L_i$, we might lost the gain of $M$. To investigate this tradeoff, the fraction $\frac{\omega|M|}{R_c+\omega L_i}$ can be used to define a new metric for M$^2$I communications, i.e., the \emph{Inductance Gain (IG)}, to characterize the benefit from metamaterial shell. We denote the inductance gain as $\mathcal{G}_{M}$, which is:
\begin{align}
\mathcal{G}_{M}=\frac{R_c^0|M^{meta}|}{(R_c^{meta}+\omega L_i^{meta})|M^0|},
\label{equ:inductance_gain}
\end{align}
where $M^{meta}$ and $L_i^{meta}$ are the inductances by using metamaterial shell, $M^0$ is the mutual inductance without using metamaterial shell. Note that without metamaterial, $L_i$ is relatively small and can be neglected here (unless in high conductive medium such as seawater and under ocean environments, which is out of the scope of this paper).

It's worth mentioning that when comparing with the coil without metamaterial shell, we set the coil's radius as $r_2$ (the outer radius of the metamaterial shell) instead of $a$ (coil radius inside the shell) for fairness. Then, we denote the resistance of the smaller coil inside the metamaterial shell as $R_c^{meta}$ and the resistance of the larger original MI coil as $R_c^0$. In the following sections, the optimization objective of the metamaterial sphere design is to maximize the inductance gain $\mathcal{G}_{M}$ in \eqref{equ:inductance_gain}.

\subsection{Modeling the Metamaterial-manipulated EM Field in M$^2$I}
According to the general framework of M$^2$I and the metamaterial enhancement strategy discussed in the previous subsection, the mutual inductance $M$ and the complex self-inductance $L_r-jL_i$ play important roles in M$^2$I communications. To quantitatively characterize the influence of the metamaterial sphere on the mutual and self-inductance, in this subsection, we investigate and model the metamaterial-manipulated electromagnetic field around both the M$^2$I transmitter and receiver. The field model is then validated by the FEM simulations. Finally, a proof-of-concept experiment is discussed to confirm the feasibility of the M$^2$I in real implementations.

It should be noted that we focus on the M$^2$I point-to-point communication in this subsection. The performance of the M$^2$I waveguide can be easily derived based on the analysis of point-to-point communication.
Moreover, the orientation of the coil inside a metamaterial shell can affect the system performance especially when two coils are perpendicular to each other. This problem can be solved by the tri-directional coil antenna \cite{Guo_oil_2014} that mounts three concentric and orthogonal signal coils together in both the transmitter and receiver. As each of the three concentric coils covers one direction in the Cartesian coordinate, the entire 3D space is covered. Then, no matter how the transmitter or receiver moves and rotates during deployment or operation, reliable wireless channel can be maintained. The tri-directional coil structure can be easily inserted into the metamaterial sphere in M$^2$I discussed in this paper. The performance is also easy to model by simply adding the fields from the three coils. However, such analysis would add unnecessary complexity and defocus the key point in this paper. Hence, we only consider coaxially-placed coils in the following analysis. Detailed discussion on the MI tri-directional coil antennas can be found in \cite{Guo_oil_2014, Tan_testbed_2015, Guo_underwater_2015}.

%It should be noted that in this paper we focus on the metamaterial enhancement to the original MI and MI waveguide communications. The MI coil considered here is a single-turn coil. All the coils in the transmitter, receiver, and relays are also coaxially placed. There are several variations of this basic setting of MI coils. For instance, the channel path loss can be reduced by simply increasing the number of turns, which is easy to model but trivial to investigate. {\color{red}here I move the orientation discussion to the end of this section.}

\subsubsection{EM Field around M$^2$I Antenna}
Consider the M$^2$I point-to-point communication between a M$^2$I transmitter and a M$^2$I receiver, as shown in Fig.~\ref{fig:sys}. The coil is located at the center of the metamaterial shell. We define the space inside and outside the shell as the first and third layer, and the shell itself is the second layer. In the first and second layers, there are standing waves while the third layer has traveling wave. Hence, spherical Bessel function of the first kind and spherical Neumann function are used to construct the solution in the first two layers. Due to the singularity of spherical Newmann function, only spherical Bessel function of the first kind is used in the first layer. Spherical Hankel function is utilized in the third layer. For the $i^{th}$ layer, the wavenumber $k_i=\sqrt{\omega^2\mu_{i}\epsilon_{i}}$, where $\epsilon_{i}=\epsilon_0 \epsilon_{ri}-j\frac{\sigma}{\omega}$, $\epsilon_0$ is the vacuum permittivity, $\epsilon_{ri}$ is the relative permittivity, $\mu_{i}=\mu_0 \mu_{ri}$, $\mu_0$ is the vacuum permeability, $\mu_{ri}$ is the relative permeability. Also, $a$ is the antenna radius, $I_0$ is the antenna current, $r$ is the distance from the origin, $\eta$ is the wave impedance, and ${\bf h}$ and ${\bf e}$ stand for magnetic field and electric field, respectively. The time-dependance $e^{j\omega t}$ is assumed. In lossy medium, the conductivity $\sigma$ has significant impact on the M$^2$I communication.

Following classical electromagnetic theory, we first construct the general solution to wave equations in spherical coordination. Then based on boundary conditions, we can obtain the complete solution. Notice that, since the magnetic dipole's radius and its wire radius are much smaller than the wavelength, the antenna only radiates TE$_{01}$. Moreover, the metamaterial shell is much smaller than the wavelength, we only need to consider the first order mode \cite{tsang2000scattering}. Thus, the unknown magnetic fields in each layer can be expressed as
\begin{subequations}
\label{equ:layer_equation}
\begin{alignat}{1}
\label{equ:firstlayer}
\hspace{-4mm} 1^{st}~layer:
\begin{cases}
{\bf h}_{r1}=\frac{-2 j \cos\theta}{\omega \mu_1 r}\alpha_1 j_1(k_1 r){\hat r},\\
{\bf h}_{\theta1}=\frac{j  \sin\theta }{\omega \mu_1 r}\alpha_1\left[j_1(k_1 r) +k_1r{j_1}'(k_1 r)\right]{\hat \theta};
\end{cases}
\end{alignat}
\vspace{-2mm}
\begin{alignat}{1}
\label{equ:layer}
\hspace{-2.4mm} 2^{nd}~layer:
\begin{cases}
{\bf h}_{r2}=\frac{-2j\cos\theta}{\omega \mu_2 r}\left[\alpha_2 j_1(k_2 r)+\alpha_3 y_1(k_2 r)\right]{\hat r},\\
{\bf h}_{\theta2}=\frac{j \sin\theta }{\omega \mu_2 r}\left\{\alpha_2 \left[j_1(k_2 r)+k_2r{j_1}'(k_2 r)\right]\right.\\
\hspace{7mm} \left.+\alpha_3\left[y_1(k_2 r)+k_2 r {y_1}'(k_2r)\right]\right\}{\hat \theta};
\end{cases}
\end{alignat}
\vspace{-1mm}
\begin{alignat}{1}
\label{equ:thirdlayer}
3^{rd}~layer:
\begin{cases}
{\bf h}_{r3}=\frac{-2 j \cos\theta}{\omega \mu_3 r}\alpha_4 {h_1^{(2)}}(k_3 r){\hat r},\\
{\bf h}_{\theta3}=\frac{j  \sin\theta }{\omega \mu_3 r}\alpha_4\left[{h_1^{(2)}}(k_3 r) +k_3r{h_1^{(2)}}'(k_3 r)\right]{\hat \theta};
\end{cases}
\end{alignat}
\end{subequations}
where $\alpha_i$ is the unknown coefficient; $j_1(kr)$ is spherical Bessel function of the first kind and order 1, and $y_1(kr)$ is spherical Neumann function of order 1, $h_1^{(2)}(kr)$ is spherical Hankel function of the second kind and order 1, and the prime symbol denotes derivative. %{\color{red} It should be noted that the dominant field is from the M$^2$I transmitter while the re-radiated field from the M$^2$I receiver is negligible, since the size of the receiver (in centimeter scale) is much smaller than the wavelength (tens of meters) and the communication range (tens of meters).}

According to Maxwell equations, the normal component of the magnetic flux (${\bf B}$) and the tangential component of the magnetic field (${\bf h}$) should be continuous at the boundary. Then by adding the excitation source and rearranging \eqref{equ:layer_equation}, we can obtain the unknown coefficients by
\begin{align}
\label{equ:matrix_trans}
{\bf A}_t={\bf S}_{meta}^{-1} {\bf \Psi}_t ,
\end{align}
where ${\bf A}_t^t=\left[\alpha_1, \alpha_2, \alpha_3, \alpha_4\right]$; ${\bf S}_{meta}$ is a coefficient matrix and ${\bf \Psi}_t$ is the excitation vector. The detailed expressions for ${\bf S}_{meta}$ and ${\bf \Psi}_t$ are provided in Appendix A. After solving \eqref{equ:matrix_trans}, by substituting the unknown coefficients $\alpha_1$, $\alpha_2$, $\alpha_3$, and $\alpha_4$ into \eqref{equ:layer_equation}, the intensity of magnetic field around the M$^2$I transmitter (outside the metamaterial sphere), i.e., ${\bf h}_{r3}$ and ${\bf h}_{\theta3}$ in \eqref{equ:thirdlayer}, can be derived. The magnetic field intensity around the M$^2$I coil (inside the metamaterial sphere), i.e., ${\bf h}_{r1}$ and ${\bf h}_{\theta1}$ in \eqref{equ:firstlayer}, can also be derived.

Similar as the transmitter, the magnetic field in each layer of the receiver can be expressed by \eqref{equ:layer_equation}. The difference is that the magnetic field in the third layer is the scattered field from the receiving shell. In order to distinguish the transmitter and receiver, the unknown coefficients are $\beta_i$ for each layer at receiver side. By substituting $\alpha_i$ with $\beta_i$ in \eqref{equ:layer_equation}, we can obtain the magnetic field intensity inside the receiver's shell.
Based on boundary conditions, $\beta_i$ can be determined by
\begin{align}
\label{equ:matrix_receive}
{\bf A}_r={\bf S}_{meta}^{-1}{\bf \Psi}_r ,
\end{align}
where ${\bf A}_r^t=\left[\beta_1, \beta_2, \beta_3, \beta_4\right]$; and similarly  detailed expressions for ${\bf S}_{meta}$ and ${\bf \Psi}_r$ are provided in Appendix A.

Once the solutions of the coefficients $\beta_1$, $\beta_2$, $\beta_3$, and $\beta_4$ are derived, the magnetic field intensity distribution around the receiver (inside the metamaterial sphere) can be expressed in the same format given in \eqref{equ:layer_equation} (replacing $\alpha_1$ with $\beta_1$).

\begin{figure}[t]
  \centering
    \includegraphics[width=0.3\textwidth]{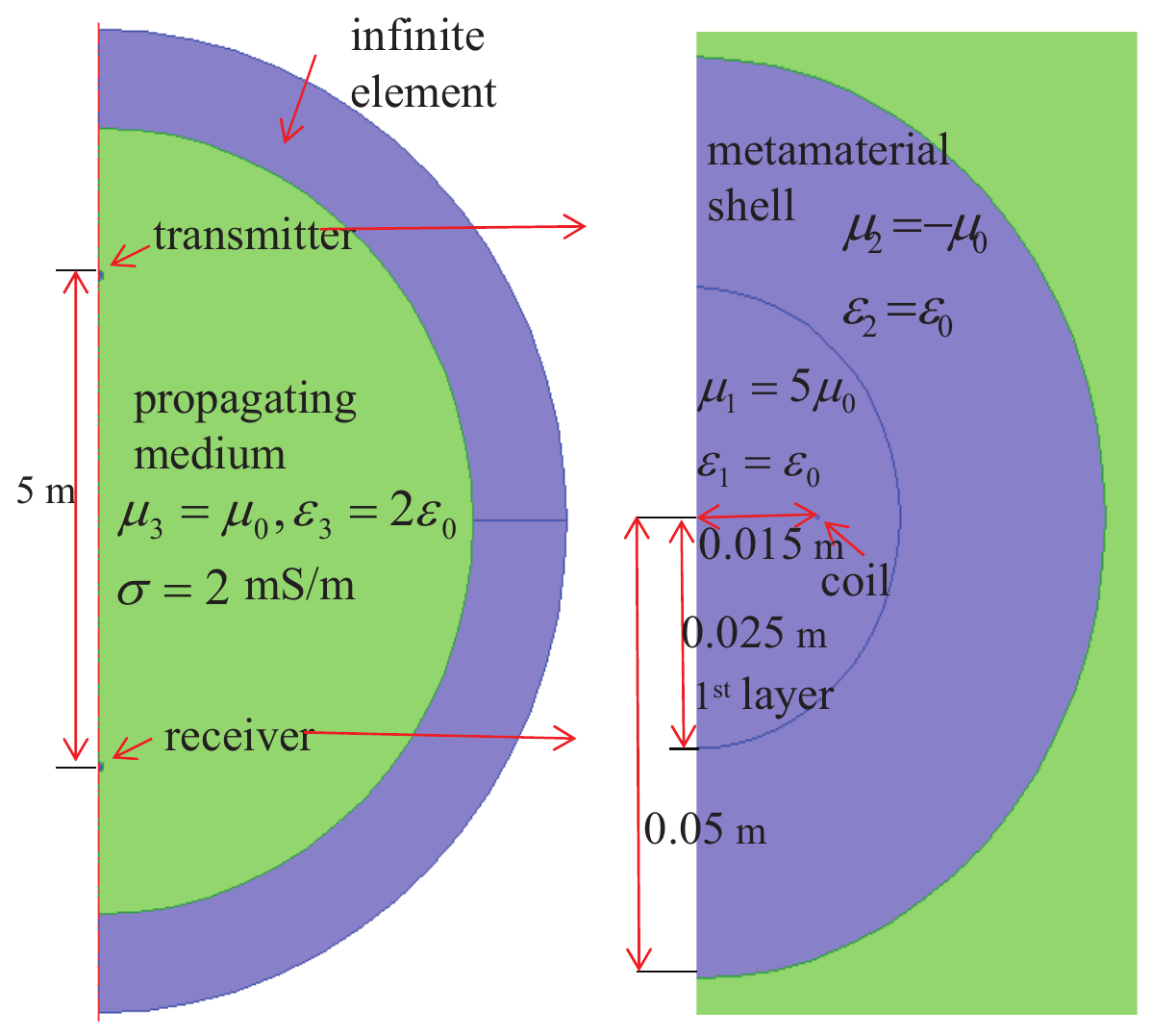}
    \vspace{-10pt}
  \caption{2D simulation model in Comsol.(right side is zoom-in of metamaterial shell enclosed coil; infinite element is utilized to extend the simulation domain toward infinity.)}
  \vspace{-8pt}
  \label{fig:comsol_model}
\end{figure}
\begin{figure}[t]
  \centering
    \includegraphics[width=0.34\textwidth]{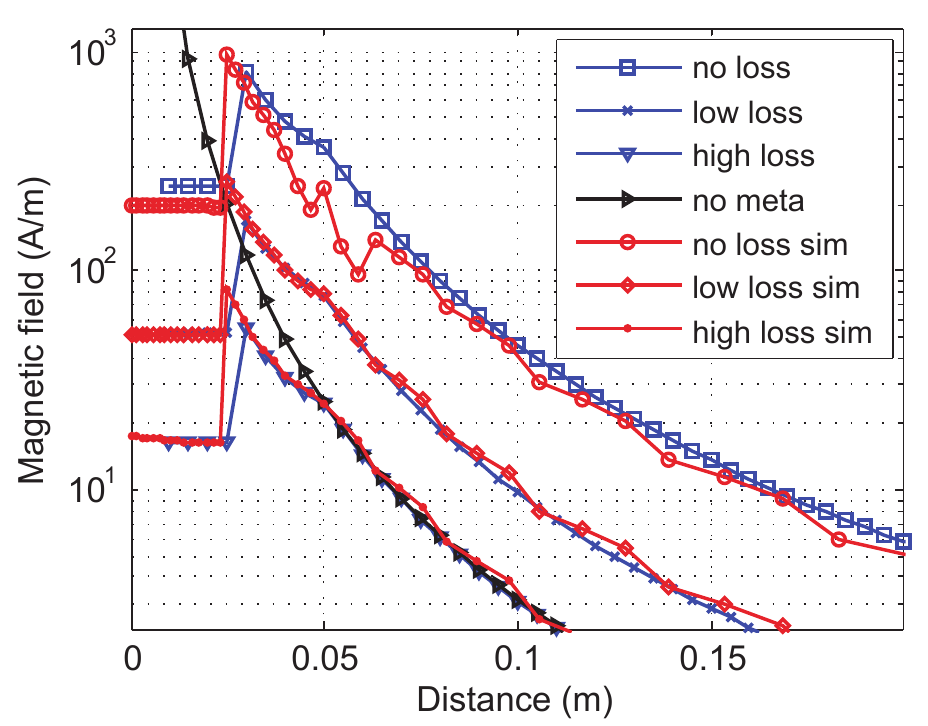}
    \vspace{-10pt}
  \caption{Magnetic field intensity around transmitting coil in soil medium.(The center of the transmitting coil is located at 0~m, extending towards receiver from 0~m to 0.2~m.)}
  \vspace{-8pt}
  \label{fig:txfield}
\end{figure}
\begin{figure}[t]
  \centering
    \includegraphics[width=0.34\textwidth]{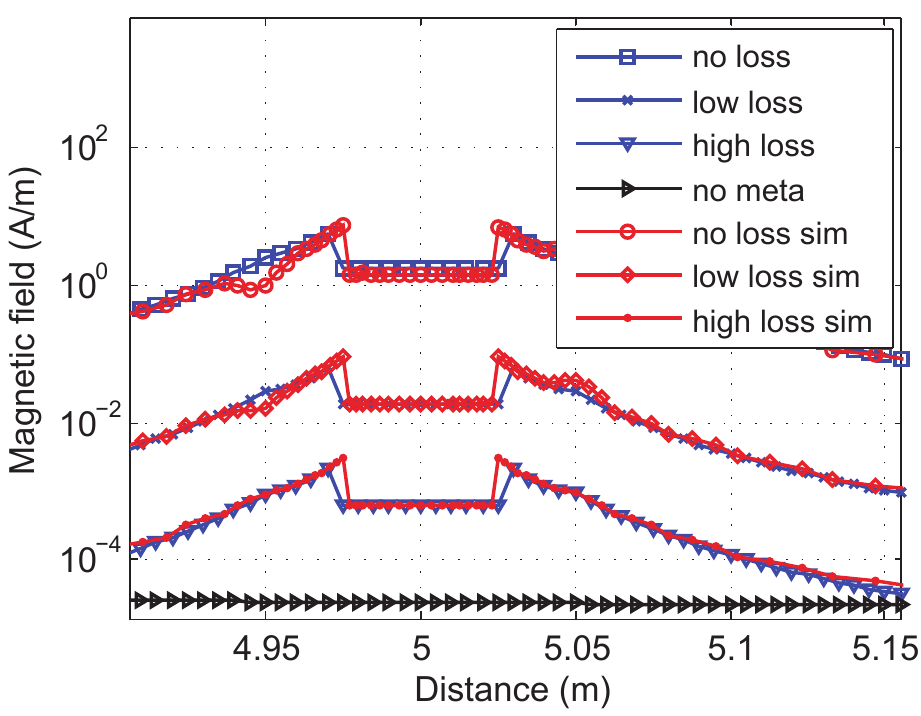}
    \vspace{-10pt}
  \caption{Magnetic field intensity around receiving coil in soil medium.(The center of the receiving coil is located at 5~m, extending from 4.9~m to 5.15~m.)}
  \vspace{-5pt}
  \label{fig:rxfield}
\end{figure}

%\begin{figure*}[t]
%\centering
%\begin{minipage}{0.3\textwidth}
%	\centering
%	    \includegraphics[width=1.75in]{fig/sim_system_1}
%	    	  \vspace{-9pt}
%	  \caption{2D simulation model in Comsol.(right side is zoom-in of metamaterial shell enclosed coil; infinite element is utilized to extend the simulation domain toward infinity.)}
%	  \label{fig:comsol_model}
%\end{minipage}
%\quad\quad
%\begin{minipage}{0.3\textwidth}
%\centering
%%\vspace{20pt}
%\includegraphics[width=1.9in]{fig/tX_field_revise}
%    \vspace{-9pt}
%  \caption{Magnetic field intensity around transmitting coil in soil medium.(The center of the transmitting coil is located at 0 m, extending towards receiver from 0 m to 0.2 m.)}
%  \label{fig:txfield}
%\end{minipage}
%\quad\quad
%\begin{minipage}{0.3\textwidth}
%\centering
%%\vspace{20pt}
%\includegraphics[width=1.9in]{fig/rtX_field_revise}
%    \vspace{-9pt}
%  \caption{Magnetic field intensity around receiving coil in soil medium.(The center of the receiving coil is located at 5 m, extending from 4.9 m to 5.15 m.)}
%  \label{fig:rxfield}
%\end{minipage}
%\vspace{-5pt}
%\end{figure*}

\subsubsection{Validation using FEM Simulations}
In this subsection, we verify the developed field model via FEM simulation in Comsol Multiphysics \cite{comsol}. The system configuration is set as follows. The overall size of the metamaterial sphere is set to be 10 cm in diameter, i.e., $r_2=0.05$ m. Such antenna size can be fit in many wireless devices. Similar to most MI communication systems, the operating frequency is set at 10 MHz. A single negative metamaterial layer is used. Without loss of generality, we set the relative permeability of the metamaterial to be $-1$, i.e., $\mu_{2}=-\mu_0$. The transmission medium outside the sphere is considered to be soil with the relative permeability $\mu_3=\mu_0$, permittivity $\epsilon_3=2\epsilon_0$, and conductivity $2$ mS/m.

The maximum inductance gain (IG) can be obtained by finding the optimal thickness of the metamaterial layer, i.e., $r_1$, and the permeability of the infilling inside the sphere, i.e., $\mu_1$ (See details in Section III-C). Since we only need to validate the field model derived in this subsection, we directly use the optimal values: $r_1=0.025$~m and $\mu_1=5\mu_0$. In addition, the size of the MI coil $a$ is supposed to be as large as possible \cite{Sun_MI_TAP_2010}. Theoretically we can set $a=r_1$. However, as $a$ approaches $r_1$, it will cause strong effect on the boundary. As suggested in \cite{Ziolkowski_chu}, $\frac{a}{r_1}=60\%$. Thus, we set $a=0.015$~m. For fair comparison, the radius of the coil without metamaterial shell is set to be $5$~cm (the same as $r_2$). It should be noted that the above parameters of the metamaterial sphere are practical. As demonstrated in \cite{PhysRevB.85.201104, Scarborough_exp}, it is possible to fabricate low loss metamaterial with unit size less than $\frac{1}{2000}\lambda$ at around 10~MHz. The metamaterial sphere thickness used in this paper ($r_1$=0.025~m) is $\frac{1}{720}\lambda$, which is well above the threshold.

%In order to match with the simulation in Comsol, the coil resistance $R_c$ is measured from the software.

%Reader may wonder that: is it fair that we set $\mu_1=5\mu_0$ for a coil with metamaterial shell to compare with a coil who does not have such a shell and $\mu=\mu_0$ everywhere. In other words, will this higher permeability inside the shell make the comparison unfair. Note that the coil without shell has the radius 0.05m and the metamaterial shell enclosed coil has radius 0.015m. Hence the area of the latter is $\frac{1}{11}$ of the former even the permeability is 5 times larger. Since the mutual inductance is almost a linear function of permeability and coil's area, the coil in metamaterial shell does not have any advantage if we do not consider negative permeability.

It should be noted that the intrinsic loss effect in metamaterial is also considered. As reported in \cite{Wang_WPT}, the measured loss is 0.05$\mu_0$. In this paper, we consider $\mu_2$ has three levels of losses, i.e., high loss, low loss, and no loss.  The corresponding parameters are: high loss $\mu_2=(-1-0.05j)\mu_0$, low loss $\mu_2=(-1-0.005j)\mu_0$ and no loss $\mu_2=-\mu_0$. Comsol simulation model is shown in Fig. \ref{fig:comsol_model}. It's an axis-symmetric model where the coordinate is cylindrical. AC/DC module is utilized here and the distance between two coils is 5~m. All the parameters in the simulations are summarized in Table \ref{table_parameters}. Different from \cite{Ziolkowski_electric}, when comparing the performances, we consider both the M$^2$I antenna and the original MI antenna have the same transmission power $P_{t}$ instead of the same antenna current. As discussed in Section III-A, the M$^2$I coil has additional frequency-dependent resistance from the imaginary self-inductance, which can consume significant power. For the considered four scenarios, the input impedance highly depends on the additional resistance ($\omega L_i$). Therefore, to make the comparison fair, we set the transmission power $P_t$ in \eqref{equ:p2p_approx} as 1 W for all of the four scenarios.

%in order to guarantee the same power (1 W) for fair evaluation, the input current for the no loss, low loss, and high loss M$^2$I antenna, as well as the original MI antenna are 17.3 mA, 70 mA, 215 mA, and 2526 mA, respectively.}

\begin{table}%[!t]
\centering
\renewcommand{\arraystretch}{1.3}
\caption{Simulation Parameters in Lossy Soil Medium}
\label{table_parameters}
\begin{threeparttable}
    \centering
\begin{tabular}{c c|c c|c c}
\hline
$\mu_1$ & 5$\mu_0$ &$\mu_3$ &$\mu_0$&$f$&10~MHz\\
\hline
$\epsilon_1$& $\epsilon_0$& $\epsilon_2$& $\epsilon_0$&$\epsilon_3$&$2\epsilon_0$\\
\hline
$r_1$&0.025~m & $r_2$&0.05~m&a&0.015~m\\
\hline
$R_c$&0.047~$\Omega $& $\sigma$&2~mS/m &$P_{t}$&1~W \\
\hline
${\mu_{r2}}^{no}$&-1&${\mu_{r2}}^{low}$&-1-0.005j&${\mu_{r2}}^{high}$&-1-0.05j\\
\hline
\end{tabular}
\begin{tablenotes}
            \item $\mu_{r2}$ is the relative permeability in the second layer.
        \end{tablenotes}
     \end{threeparttable}
     \vspace{-5pt}
\end{table}

\begin{figure}%[H]
  \centering
    \includegraphics[width=0.32\textwidth]{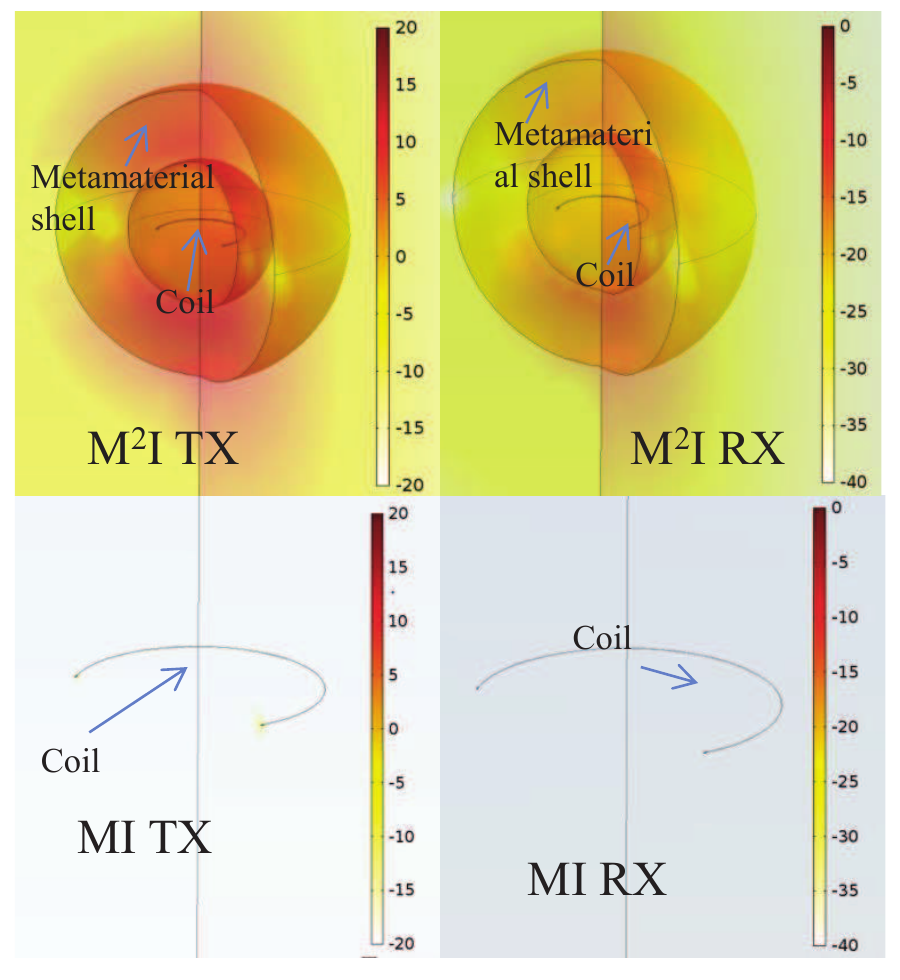}
    \vspace{-10pt}
  \caption{Magnetic field (Unit: A/m) of M$^2$I communication (upper) and conventional MI communication(lower) in dB scale. Distance is 5~m and metamaterial shell has no loss.}
  \vspace{-5pt}
  \label{fig:revised_p2psimulation}
\end{figure}

As shown in Fig. \ref{fig:txfield} and Fig. \ref{fig:rxfield}, the magnetic field intensity derived by the theoretical field model has good match with the FEM simulation results at both transmitter and receiver side.
On the one hand, we observe that by using the metamaterial shell with optimal parameters, the magnetic field intensity can be increased by more than 1 order of magnitude, compared with the original MI system. On the other hand, the metamaterial loss can dramatically reduce the gain brought by the metamaterial sphere. Also, notice that the receiver side has larger gain than transmitter side. The reason is the magnetic field is amplified again by the receiver's metamaterial shell.
In Fig. \ref{fig:revised_p2psimulation}, the enhancement on the magnetic field by no loss M$^2$I is visually shown by the Comsol simulation. The two configurations have the same input power. The coil with metamaterial shell can generate much stronger magnetic field. Also, the amplification at the receiver side is obvious. In contrast, the original MI receiving coil has very weak field which almost cannot be seen from the figure.
%Since the objective here is to verify the theoretical model, more detailed discussion on the M$^2$I performance is provided in Section III.C and Section IV.
\begin{figure}[t]
  \centering
    \includegraphics[width=0.42\textwidth]{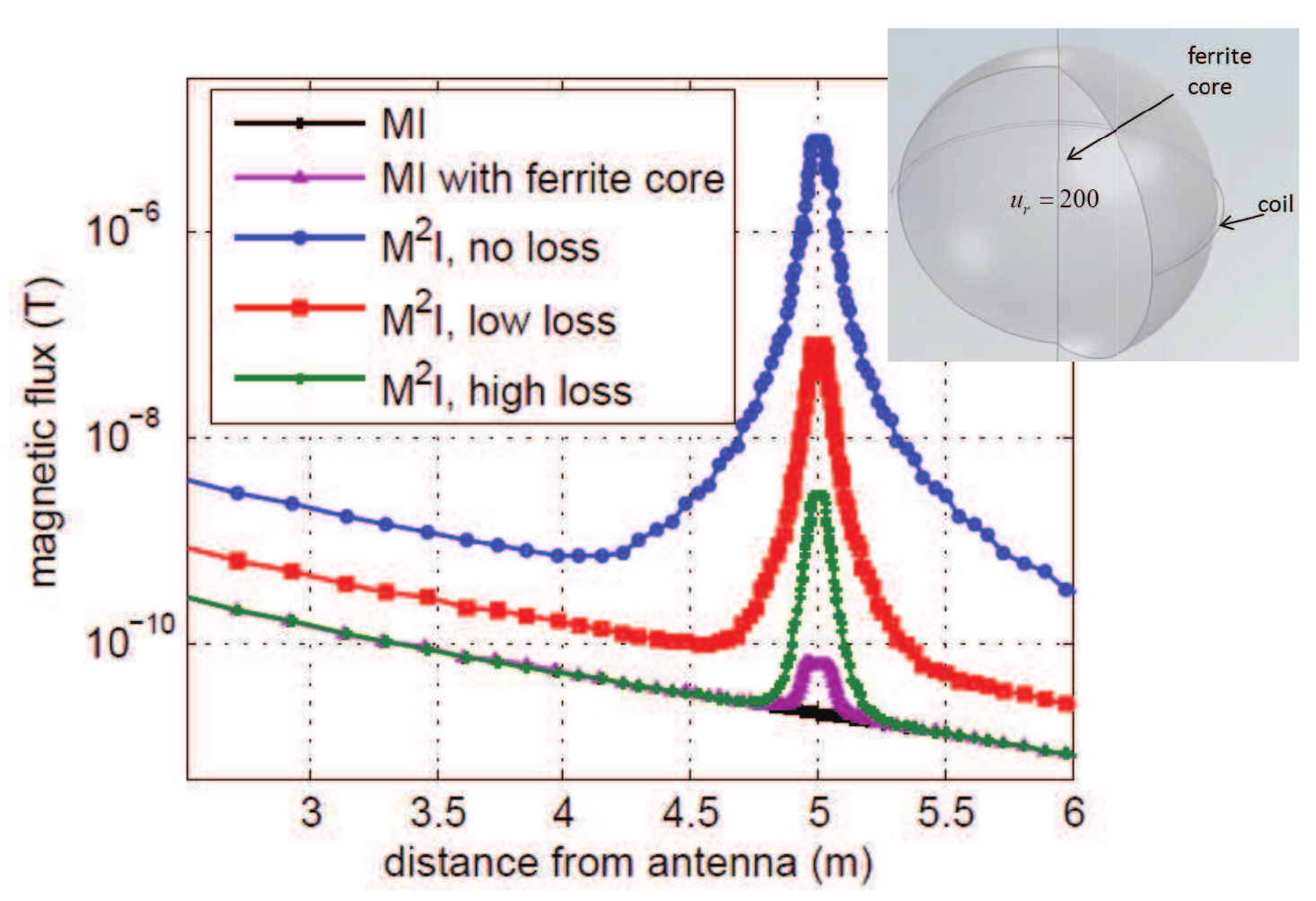}
    \vspace{-10pt}
  \caption{FEM simulation of magnetic flux intensity (T). The center of the receiving coil is located at 5~m, extending from 2.5~m to 6~m.}
  \vspace{-5pt}
  \label{fig:ferrite}
\end{figure}

In addition, the effect of a ferrite core at the receiver is discussed since this strategy is widely used to improve MI's performance. As shown in Fig. \ref{fig:ferrite}, we consider that there is a spherical core inside the receiving MI coil. The radius of the core is the same as the outer radius of metamaterial shell, i.e., 0.05~m. The relative permeability of the core is 200. Configurations of MI without ferrite core and M$^2$I are the same as previous discussions. Note that, due to the high permeability, the magnetic field (A/m) inside the core is very small. However, since the ferrite core has large permeability, we obtain a large magnetic flux intensity ($B =\mu H$). As shown in Fig. \ref{fig:ferrite}, the enhancement from the ferrite core in the original MI is much smaller than that from M$^2$I, which proofs the more significant enhancement of M$^2$I.

\begin{figure}[t]
%\begin{minipage}{0.38\textwidth}
\centering
  \subfigure{
    \includegraphics[width=0.15\textwidth]{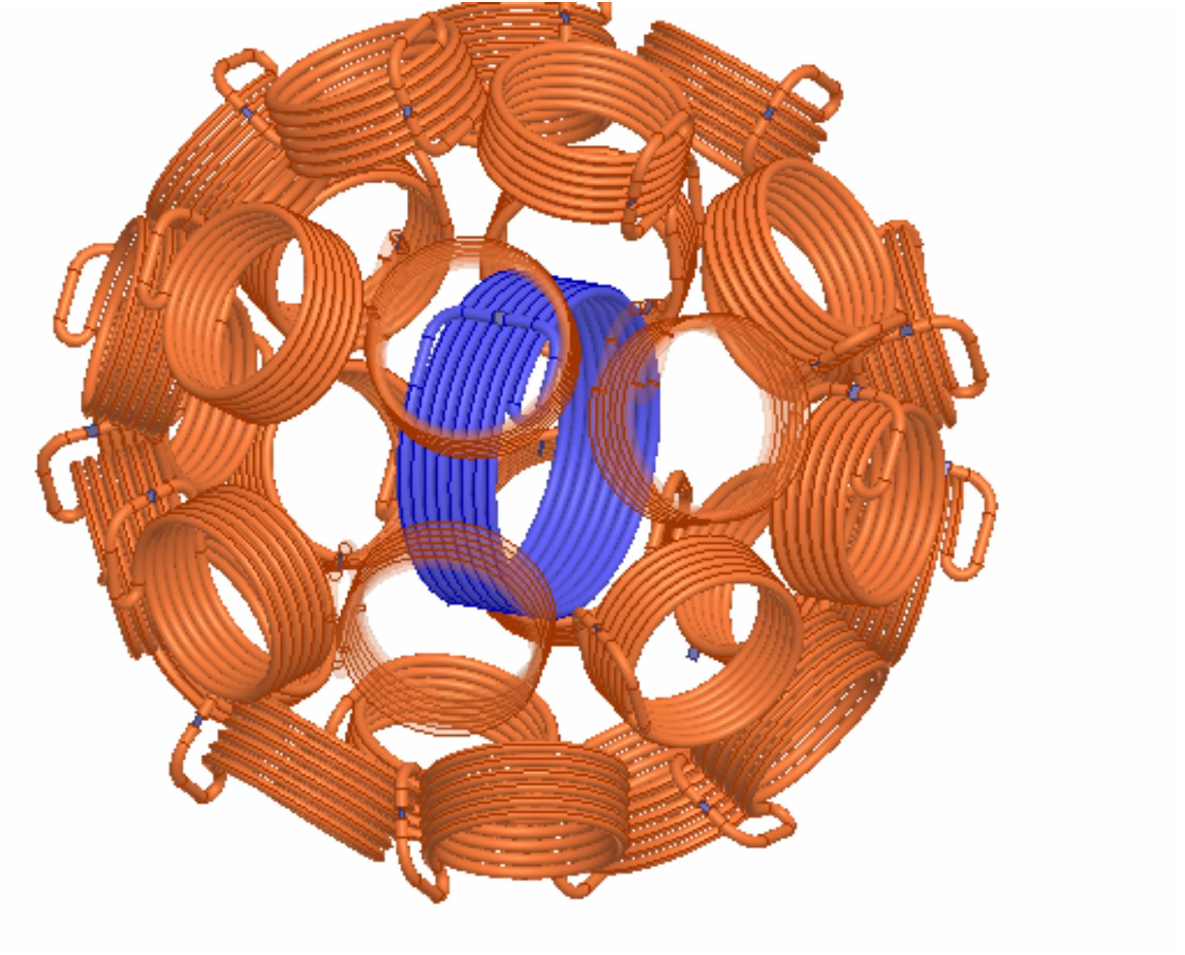}}\quad
  \subfigure{%
    \includegraphics[width=0.15\textwidth]{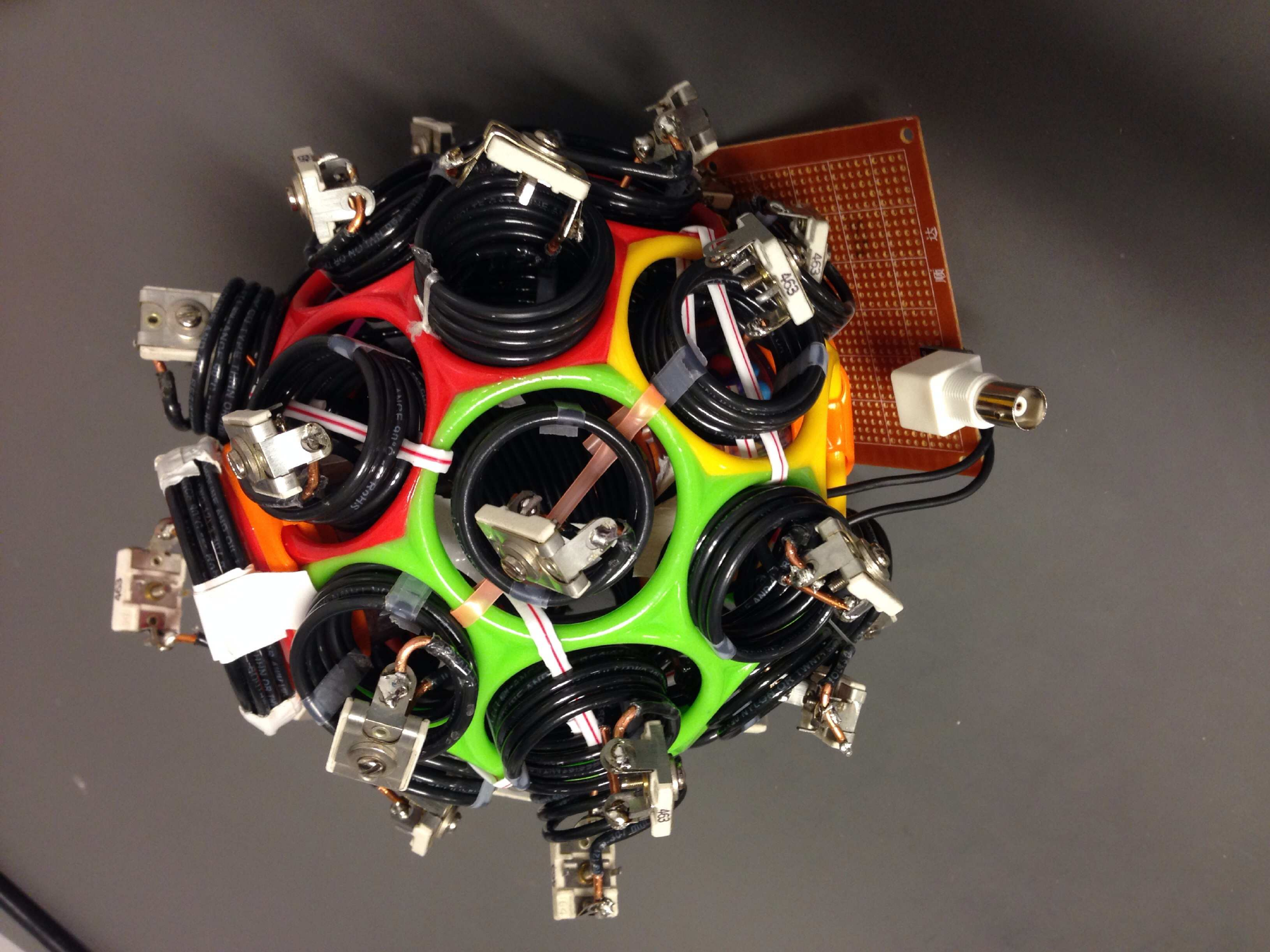}}
    \caption{The design model of the metamaterial sphere (left) and the implemented metamaterial sphere prototype (right).}
        \label{fig:sphere_model}
%\end{minipage}\quad\quad
    \vspace{-5pt}
\end{figure}

\begin{figure}[t]
\centering
\begin{minipage}{0.23\textwidth}
\centering
  \vspace{-0.3cm}
  \includegraphics[width=0.95\textwidth]{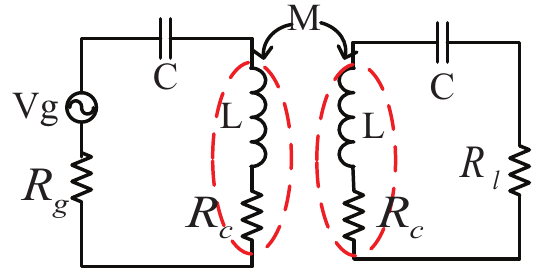}
  \vspace{-0.1cm}
  \caption{Experimental equivalent circuit.}
    \label{fig:experiment_config}
\end{minipage}\quad
\begin{minipage}{0.23\textwidth}
\centering
  \includegraphics[width=0.8\textwidth]{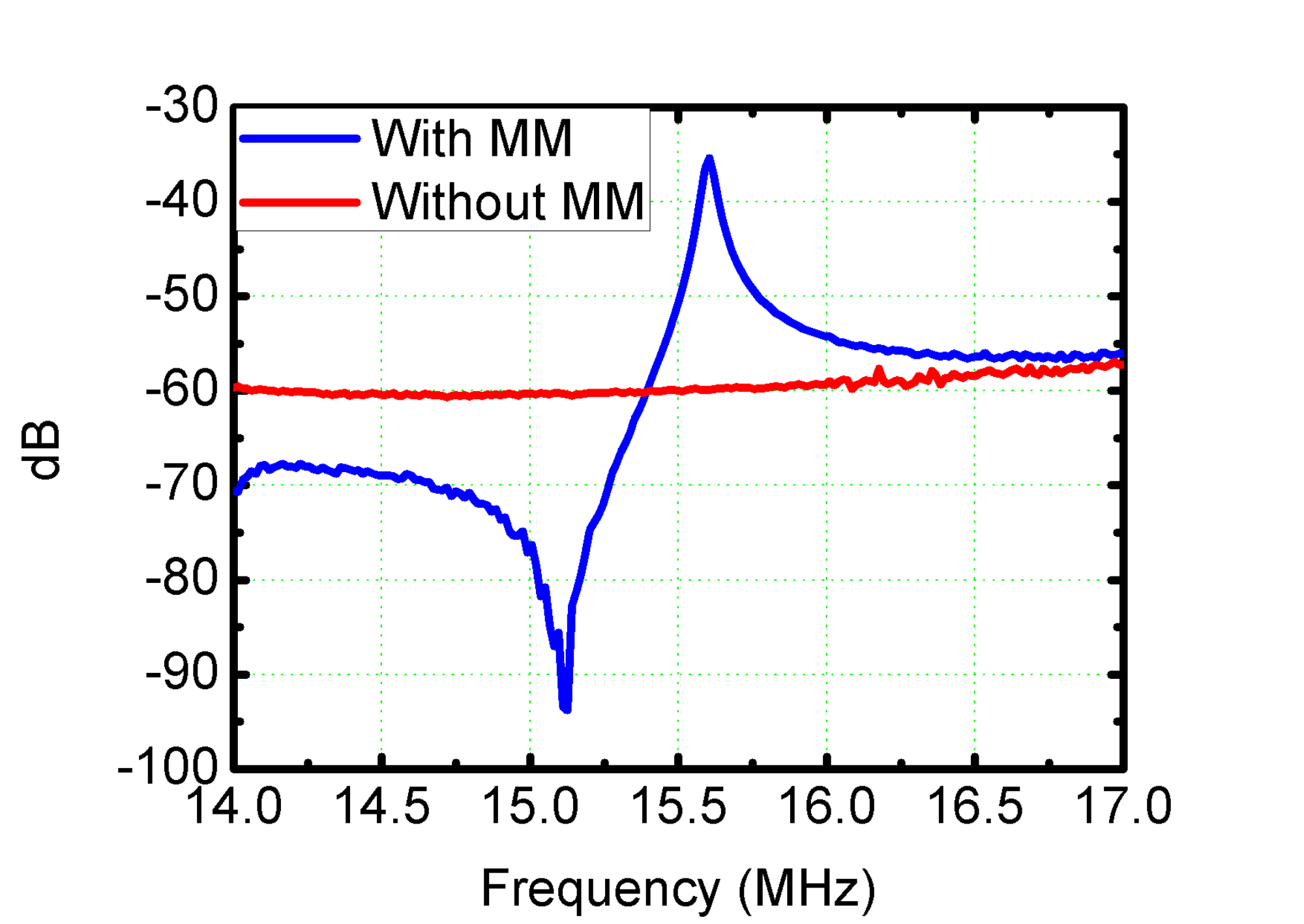}
    \vspace{-0.1cm}
  \caption{Measured $S_{21}$ parameter.}
    \label{fig:exp_10cm}
\end{minipage}
    \vspace{-15pt}
\end{figure}

\subsubsection{Experimental Validation}

To validate the predicted M$^2$I enhancement, a proof-of-concept prototype of metamaterial sphere is designed and implemented. As shown in Fig.~\ref{fig:sphere_model}, the ideal spherical shell is approximated by a 36-face polyhedron. The diameter of the polyhedron is approximately 10~cm. Each face of the polyhedron forms a metamaterial unit, which is a 6-turn coil with 1.5~cm radius and loaded with a variable capacitor. An 8-turn MI coil antenna with 2.5~cm radius is fixed in the center of the polyhedron. By tuning the variable capacitor, the fabricated metamaterial shell can achieve the resonance at 15.5~MHz (i.e., achieves the negative magnetic permeability at 15.5~MHz). The equivalent circuit for transmitter and receiver is shown in Fig. \ref{fig:experiment_config}. Thanks to the lossless environment, $L_i$ can be neglected. As a result, MI and M$^2$I have the same equivalent circuit. The inductance is still compensated by a capacitor to make the circuit resonant. The difference from the theoretical model is that the source is not ideal (it has resistance $R_g$) and also the receiver load $R_l$ is fixed.

Based on the equivalent circuit and loose coupling assumption, \eqref{equ:p2p_pathloss} for both MI and M$^2$I can be updated according to the experimental configuration:
\begin{align}
\label{equ:experiment}
\frac{P_r}{P_t}=\underbrace{\frac{R_l}{(R_g+R_c)(R_l+R_c)^2}}_{\text{T}}\cdot\omega^2|M|^2=T\cdot\omega^2|M|^2.
\end{align}
In \eqref{equ:experiment}, while $T$ is the same in both MI and M$^2$I, the only difference is $M$. Although $R_g$ and $R_l$ are fixed so that we cannot change $R_l$ to match with $R_c$ (as in the theoretical model), the performance difference between MI and M$^2$I keeps the same. Matching the resistance can only increase $T$, which is the same for MI and M$^2$I. The differences on power ratios between MI and M$^2$I are still the same, i.e., the enhancement of M$^2$I keeps the same regardless the resistance is matched or not.

The experimental setup is shown in Fig.~\ref{fig:sphere_model}. The MI transmitter is enclosed by the metamaterial shell while the MI receiver is an original MI coil antenna (8-turn, 2.5~cm radius). The Agilent 8753E RF network analyzer is used to measure the $S_{21}$ parameter of the prototype. The transmitter coil (inside the metamaterial shell) and the original MI receiver coil (the wire loop) are connected to the two ports of the network analyzer, respectively. On the one port, the network analyzer feeds the transmitter coil with 14 to 17~MHz signals. On the other port, the signal received by the MI receiver coil is input to the network analyzer to display the measurements. For comparison, we also conduct the same experiment for a transceiver pair without the metamaterial shell.

Fig.~\ref{fig:exp_10cm} gives the received signal strength of the MI communication with and without the metamaterial shell, i.e.,  M$^2$I and MI, respectively. The receiver is placed 10~cm away from the M$^2$I transmitter. Around the resonance frequency, i.e., 15.5~MHz, more than 20~dB enhancement is observed when the metamaterial shell is used, which is consistent with the theoretical and the simulation prediction. Therefore, the concept of metamaterial enhanced MI communication can be proved by this initial prototype and experiment. It should be noted that, the experiment results are not directly compared with the numerical results in this paper. The reason is that currently it is still an open issue to model a fabricated metamaterial. Hence, we cannot exactly determine the values of metamaterial thickness, effective radius, and effective permeability, which prevent a directly and apple-to-apple comparison between the experiments and the numerical results.

\subsection{Analytical M$^2$I Channel Model}

Based on the field model derived in the last subsection, the self-inductance and the mutual inductance as well as other channel parameters in the M$^2$I communication can be calculated. However, the field model requires complicated calculations, such as the inverse of the matrix of Bessel functions. Therefore, the model is limited to numerical results and can not provide analytical insights on the metamaterial enhancement mechanism, not to say the optimization of the system. To this end, we develop the analytical M$^2$I channel model with an explicit and tractable expressions of self-inductance, mutual inductance, and path loss, as well as bandwidth and capacity. Then based on the analytical model, the resonance condition as well as the optimal configuration of M$^2$I communications are investigated.

\subsubsection{Deriving Explicit Expressions for Analytical Channel Model}

We start the investigation by calculating the M$^2$I self-inductance and mutual inductance based on the developed field model.
Due to the influence of the metamaterial shell, the self-inductance consists of two parts: one is the original inductance generated by the coil and the other one is the inductance contributed by the metamaterial enhancement:
\begin{align}
\label{equ:self_induction}
L \!=\!\frac{\Phi_1}{I_0}\!=\!\frac{1}{I_0}\iint\displaylimits_{S}{\bf B}\cdot{\hat n}~dS\!\simeq\! L_0\!+\! \frac{4 \pi \alpha_1 }{j \omega k_1 I_0  }\left[1\!-\!\frac{\sin (k_1 a)}{k_1 a}\right],
\end{align}
where $\Phi_1$ is the magnetic flux through the transmit coil; $S$ is the area of the coil; ${\hat n}$ is the orientation of the coil; and $L_0$ is the coil's self-inductance without the shell, which can be approximated by $L_0=\mu_1 a [\ln(\frac{8a}{r_w})-2]$, where $r_w$ is the wire radius \cite{503178}. Similarly, the mutual inductance is the magnetic flux through the receive coil over the current in the transmit coil, which can be expressed as
\begin{align}
\label{equ:mutual_induction}
M\simeq \frac{4 \pi \beta_1 }{j \omega k_1 I_0 }\left[1-\frac{\sin (k_1 a)}{k_1 a}\right].
\end{align}
Note that, the reradiated field from the receiver coil is considered in this mutual inductance since it is bidirectional.

According to \eqref{equ:self_induction} and \eqref{equ:mutual_induction}, the key coefficients that determines the self-inductance and mutual inductance are $\alpha_1$ and $\beta_1$. To derive $\beta_1$, $\alpha_4$ is also needed to be calculated. Those coefficients need to be derived through \eqref{equ:matrix_trans} and \eqref{equ:matrix_receive}, which require the calculation of the inverse of a matrix consisting of different types of Bessel functions. To derive tractable channel model, such functions need to be simplified. Fortunately,
since the shell and the antenna are electrically small, i.e. $k_3 r_2 << 1$, $L$ and $M$ can be simplified as
\begin{align}
\label{equ:approx_L}
{\widetilde {L}}\approx \!L_0\!+\!\frac{\pi \rho_1 a^4 \mu_1 \left[r_2^3(\mu_1\!-\!\mu_2)(\mu_2\!+\!2\mu_3)\!+\!r_1^3(\mu_2\!-\!\mu_3)(2\mu_1\!+\!\mu_2)\right]}{18\rho_2(\rho_{3i}+\rho_{3r}j)^2 r_1^4r_2^4\mu_2\mu_3 \det({\bf \widetilde{S}}_{meta})},
\end{align}
\begin{align}
\label{equ:approx_m}
{\widetilde M}=\frac{-j}{2}\frac{\pi a^4 \rho_1^2 \mu_1^2 (\rho_{3i}-\rho_{3r}j)^2}{\rho_2^2 r_1^2 r_2^2 \mu_3(\rho_{3i}^2+\rho_{3r}^2)^2}\frac{\hbar(r)}{[\det({\bf \widetilde{S}}_{meta})]^2},
\end{align}
where $\rho_1=k_1$, $\rho_2=-jk_2$, $\rho_{3r}=\Re{(k_3)}$, $\rho_{3i}=-\Im{(k_3)}$, all $\{\rho_x\}$ are real positive numbers, and $\hbar(r)$ is a function of distance which is determined by the antenna pattern. Here $\det({\bf \widetilde{S}}_{meta})$ is the approximation of $\det({\bf {S}}_{meta})$, which is
\begin{align}
\label{equ:det_approx}
&\det({\bf {S}}_{meta})\approx \det({\bf \widetilde{S}}_{meta})+o(\frac{1}{{\bar k} {\bar r} })= \hfill\\
&\frac{\rho_1\left[2r_1^3(\mu_1-\mu_2)(\mu_3-\mu_2)-r_2^3(2\mu_2+\mu_1)(2\mu_3+\mu_2)\right]}{9 \rho_2(\rho_{3i}+\rho_{3r}j)^2 r_1 r_2^4 \mu_2\mu_3}+o(\frac{1}{{\bar k} {\bar r} }),\nonumber
\end{align}
where $\bar k$ is the asymptotic approximation of all the wavenumbers and $\bar r$ is the asymptotic approximation of all the radii. The detailed deductions for this approximation is provided in Appendix B.

%Note that
%\begin{align}
%\Re(k_3^2)=\rho_{3r}^2-\rho_{3i}^2.
%\end{align}
%{\color{red}{We may delete this sentence (when we need the this formula, I have provided the value),  Since the operating frequency is 10MHz and the soil medium has relatively low conductivity, $|k_3|$ is much smaller than 1 [XXXXXX  why?  XXXXXXX]}. Accordingly, $\Re(k_3^2)$ has a value close to zero and $\rho_{3r} \approx \rho_{3i}$}.

\begin{figure*}[t]
\centering
\begin{minipage}{0.31\textwidth}
	\centering
	    \includegraphics[width=2.08in]{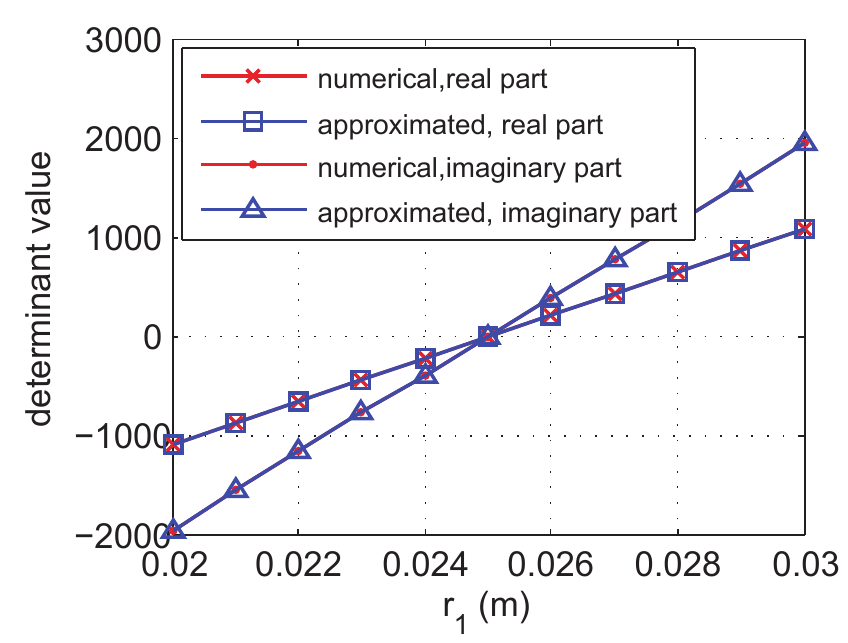}
	    	  \vspace{-8pt}
	  \caption{Comparison of $\det({\bf {S}}_{meta})$ (numerical) and $\det({\bf \widetilde{S}}_{meta})$ (approximated).}
	  \label{fig:determinant}
\end{minipage}
\quad
\begin{minipage}{0.31\textwidth}
\centering
%\vspace{20pt}
\includegraphics[width=2.03in]{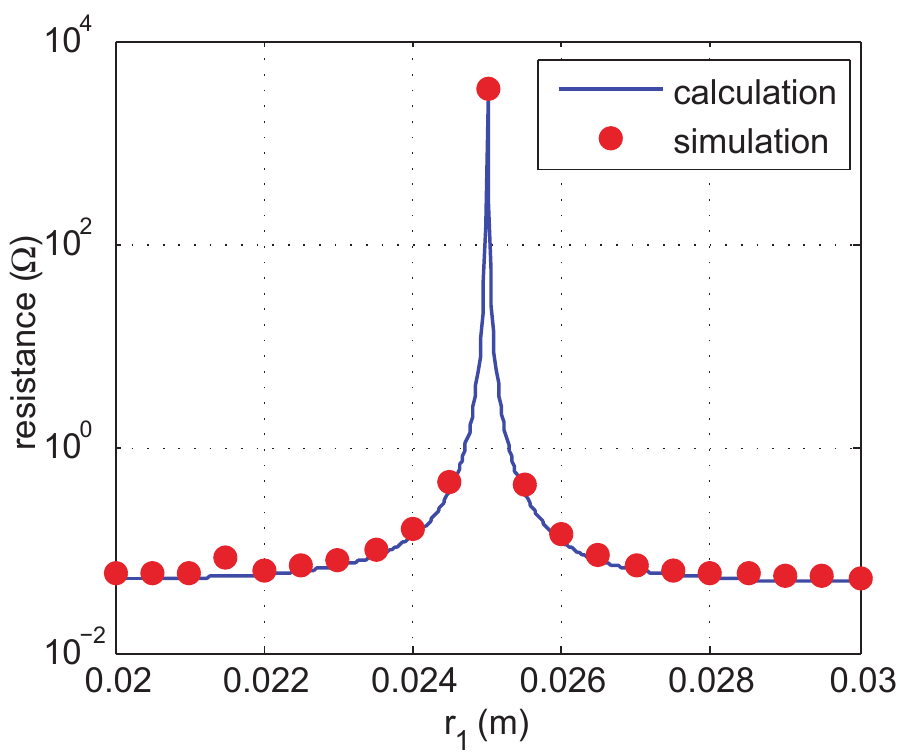}
    \vspace{-9pt}
  \caption{Resistance versus $r_1$.}
  \label{fig:3d_wo_t}
\end{minipage}
\quad
\begin{minipage}{0.31\textwidth}
	\centering
	    \includegraphics[width=2.03in]{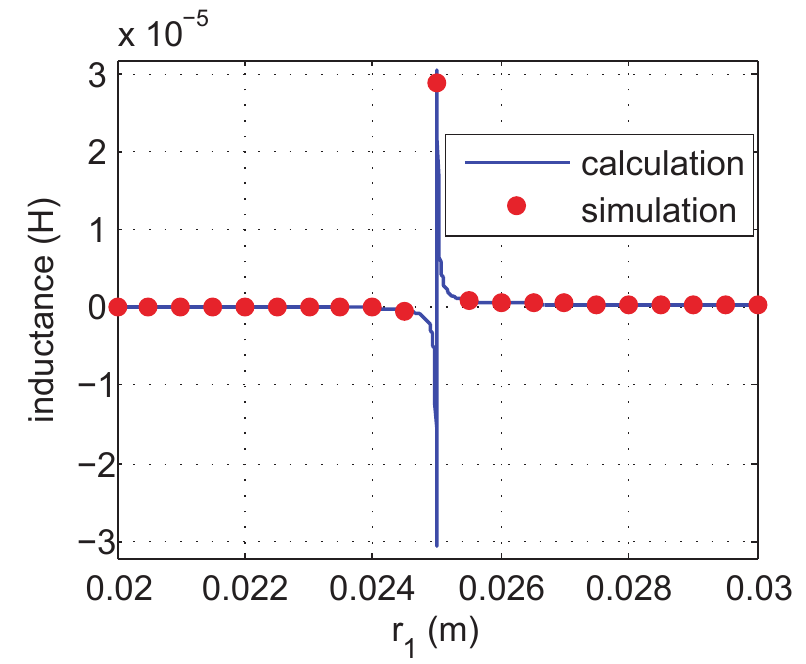}
	    	  \vspace{-12pt}
	  \caption{Inductance versus $r_1$.}
	  \label{fig:3d_wi_t}
\end{minipage}
\vspace{-7pt}
\end{figure*}

In Fig. \ref{fig:determinant}, the accuracy of the approximation $\det({\bf \widetilde{S}}_{meta})$ is numerically evaluated. The system configuration and parameters are the same as Section III-B, where the no loss case of $\mu_2$ is used. We increase $r_1$ continuously. As shown in the figure, the approximation has good match with the exact numerical results. Hence, the self- and mutual inductance in M$^2$I communications can be accurately and explicitly expressed by \eqref{equ:approx_L}, \eqref{equ:approx_m}, and \eqref{equ:det_approx}. Then the channel path loss of the M$^2$I point-to-point and waveguide communication can be derived by substituting ${\widetilde {L}}$ and ${\widetilde M}$ into \eqref{equ:p2p_approx} and \eqref{equ:wg_pathloss}. Moreover, the bandwidth and the channel capacity can also be calculated based on the path loss formula, which are discussed with the M$^2$I channel analysis in Section IV.

It should be noted that the determinant changes its sign at the resonant point ($r_1$=0.025~m). Such change causes the negative self-inductance in M$^2$I, which has not been observed in existing works. We will discuss this unique property in next subsection.

\subsubsection{Optimal Configuration of M$^2$I Communications}
The transmitter and receiver are connected by mutual inductance $M$. According to \eqref{equ:approx_m}, $M$ can be maximized if $\det({\bf \widetilde{S}}_{meta})$ is zero (i.e., $\det({\bf {S}}_{meta})$ is very small). There exists an optimal metamaterial sphere thickness $r_1$ that can greatly reduce the value of $\det({\bf {S}}_{meta})$. As a result, the mutual inductance $M$ can be significantly increased. Then the magnetic field intensity around both M$^2$I transmitter and receiver can be dramatically enhanced. The condition to achieve such enhanced peak is to find $r_1$ that makes $\det({\bf \widetilde{S}}_{meta})=0$. The solution to $\det({\bf \widetilde{S}}_{meta})=0$ can be developed as
\begin{align}
\label{equ:resonnace}
\frac{r_1}{r_2}=\sqrt[3]{\frac{(2\mu_3+\mu_2)(2\mu_2+\mu_1)}{2(\mu_2-\mu_3)(\mu_2-\mu_1)}}.
\end{align}

If the metamaterial sphere satisfies \eqref{equ:resonnace}, it achieves the metamaterial resonance. Such resonance cannot be achieved if $\mu_2$ is positive since $r_1<r_2$, which necessitates the usage of the metamaterials as the second layer of the sphere. Also, if we use a ferrite core for coil antenna to improve the performance, $\mu_1$ can be increased to more than 100$\mu_0$. By adjusting $r_1$ we can still find the resonance configuration.

This resonance condition is also observed in other metamaterial antenna designs \cite{Ziolkowski_electric} and \cite{Engheta_Meta}, where it appears that the resonance is the optimal operating mode since it amplifies the radiated power in the far field to the maximum extent. Any deviation from the resonance can significantly deteriorate the antenna's performance. In contrast, the M$^2$I communications depend on the near field where the radiated power is not as important as in the far field communications. Moreover, in lossy media, the resonance not only maximizes the mutual inductance $M$ but also maximizes the frequency-dependent resistance $\omega L_i$. Therefore, the role of metamaterial resonance in M$^2$I communications needs a major reexamination.

The frequency-dependent resistance $\omega L_i$ comes from the imaginary part of the self-inductance $L$. Hence, we investigate the effect of metamaterial resonance on the self-inductance $L$ given in \eqref{equ:approx_L}.
Under the resonant condition in \eqref{equ:resonnace}, ${\tilde L}$ can be updated as
\begin{align}
\label{equ:alpha_resonant}
{\tilde L}^{rc}\!\approx\! L_0\!+\! \frac{\rho_1 \pi a^4 \mu_1 \left[r_2^3(\mu_1\!-\!\mu_2)(\mu_2\!+\!2\mu_3)\!+\!r_1^3(\mu_2\!-\!\mu_3)(2\mu_1\!+\!\mu_2)\right]}{18 \rho_2(\rho_{3i}\!+\!\rho_{3r}j)^2 r_1^4r_2^4\mu_2\mu_3 \cdot o(\frac{1}{{\bar k}{\bar r}})}.
\end{align}
In \eqref{equ:alpha_resonant}, the absolute value of the second term is maximized since $o(\frac{1}{{\bar k}{\bar r}})$ is the minimum value of $\det({\bf {S}}_{meta})$. If the transmission medium is lossless, such as the air medium in most existing works, the second term in \eqref{equ:alpha_resonant} is real since the wavenumber of the medium $k_3$ is real (i.e., $\rho_{3i}=0$). Therefore, the self-inductance $L$ is real, which can be compensated by the capacitor. Hence, even the self-inductance is maximized, no additional loss is introduced to the M$^2$I system. However, in the lossy medium considered in this paper, the wavenumber $k_3$ becomes complex. Consequently, the imaginary part of $L$ (i.e., the frequency-dependent resistance) in \eqref{equ:alpha_resonant} is also maximized when resonant, which causes significant loss in M$^2$I.

Fig. \ref{fig:3d_wo_t} shows the total resistance ($R_{coil}^{total}=R_c+\omega L_i$) of a M$^2$I coil in the soil medium as a function of the sphere thickness $r_1$, based on both the developed model and FEM simulations. As predicted, the coil resistance is extremely large when the sphere is resonant ($r_1=0.025$ m). Hence, in M$^2$I, the resonance condition amplify both the mutual inductance $M$ and $\omega L_i$. As the sphere thickness $r_1$ moves away from the resonant condition, $L_i$ approximates to 0. As a result, the frequency-dependent resistance disappears and only the coil wire resistance is left, i.e., $R_{coil}^{total}\approx R_c$. Fig. \ref{fig:3d_wi_t} shows the calculated and simulated inductance (i.e., the real part of $L$). Similarly to the imaginary part in Fig. \ref{fig:3d_wo_t}, the real part of $L$ is dramatically amplified at the resonance point.

According to \eqref{equ:inductance_gain}, the inductance gain $\mathcal{G}_M$ between the M$^2$I transceivers is in fact determined by the ratio $M/L_i$. The effect of resonance on $\mathcal{G}_M$ is not clear yet since both $M$ and $L_i$ are maximized at the resonance point. However, according to \eqref{equ:self_induction} and \eqref{equ:mutual_induction}, $M$ is inversely propositional to $[\det({\bf \widetilde{S}})]^2$ while $L$ is inversely propositional to $\det({\bf \widetilde{S}})$. Considering that the resonance condition is in fact $\det({\bf \widetilde{S}})=0$, $M$ is more significantly amplified than $L_i$ when resonant. Hence, we can conclude that the metamaterial resonance is still the optimal operation status in M$^2$I. However, due to the same resonance effect of the frequency-dependent resistance (which incurs loss), the system performance does not deteriorate as fast as existing metamaterial antennas when $r_1$ deviates away from the optimal value. Hence, the M$^2$I system is not very sensitive to the size deviations, which is favorable in practical device fabrication.

Before numerically investigating the effects of resonance on the inductance gain $\mathcal{G}_M$, we first investigate an interesting observation in Fig. \ref{fig:3d_wi_t}, where the inductance becomes negative when $r_1$ is a little smaller than 0.025 m (the resonance condition). To find out the reason of the negative self-inductance, we analyze the $L$ under the non-resonant condition.
When the resonant condition \eqref{equ:resonnace} is not satisfied, the first term in \eqref{equ:det_approx} becomes dominant. Then ${\tilde L}$ can be expressed as
\begin{align}
{\tilde L}^{nrc}\!\!&\approx\! L_0\!+\! \frac{ \pi a^4 \mu_1 }{2 r_1^3}\frac{r_2^3(\mu_1\!-\!\mu_2)(\mu_2\!+\!2\mu_3)\!+\!r_1^3(\mu_2\!-\!\mu_3)(2\mu_1\!+\!\mu_2)}{2r_1^3(\mu_1\!-\!\mu_2)(\mu_3\!-\!\mu_2)\!-\!r_2^3(2\mu_2\!+\!\mu_1)(2\mu_3\!+\!\mu_2)}\nonumber\\
&=L_0\!+\! \frac{\pi a^4 \mu_1 }{2 r_1^3}\frac{\ell_n}{\ell_d},
\label{equ:alpha_nonresonant}
\end{align}
where $\ell_n=r_2^3(\mu_1-\mu_2)(\mu_2+2\mu_3)+r_1^3(\mu_2-\mu_3)(2\mu_1+\mu_2)$ and $\ell_d=2r_1^3(\mu_1-\mu_2)(\mu_3-\mu_2)-r_2^3(2\mu_2+\mu_1)(2\mu_3+\mu_2)$. From \eqref{equ:alpha_nonresonant}, we observe that the imaginary part of $L$ disappears if the metamaterial sphere is not resonant, which is consistent with the results shown in Fig. \ref{fig:3d_wo_t} and Fig. \ref{fig:3d_wi_t}. An interesting observation is that the real ${\tilde L}^{nrc}$ can be negative.
\begin{figure}[t]
  \centering
    \includegraphics[width=0.3\textwidth]{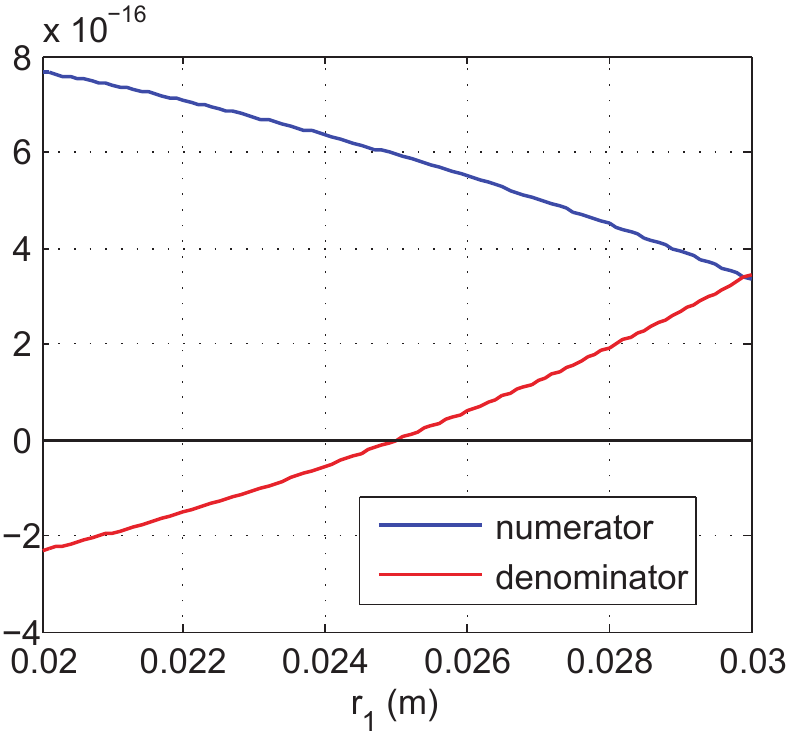}
    \vspace{-10pt}
  \caption{The numerator $\ell_n$ and denominator $\ell_d$ in \eqref{equ:alpha_nonresonant} as function of the sphere thickness $r_1$. }
  \vspace{-12pt}
  \label{fig:sign}
\end{figure}

\begin{figure}[t]
\centering
%\vspace{20pt}
%\subfigure[]{
%    \includegraphics[width=1.45in]{fig/sign}
%    \label{fig:sign}}\quad
\subfigure[]{
    \includegraphics[width=1in]{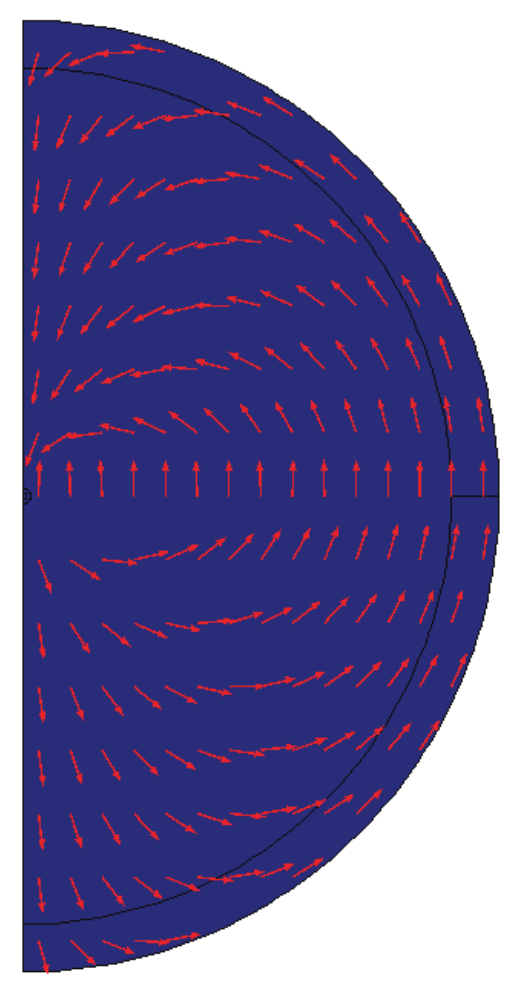}
    \label{fig:negative_inductance}}
  \subfigure[]{
    \includegraphics[width=1in]{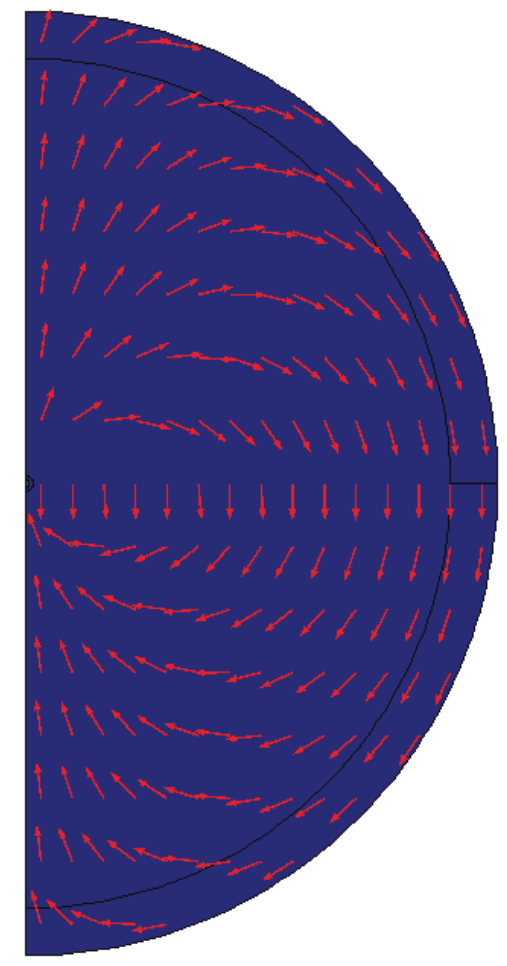}
    \label{fig:positive_inductance}}
  \vspace{-5pt}
  \caption{(a) Direction of magnetic field when $r_1=0.024$~m. (b) Direction of magnetic field when $r_1=0.026$~m. }
  \vspace{-8pt}
  \label{fig:field_direction}
\end{figure}

Fig. \ref{fig:sign} shows the value of the numerator $\ell_n$ and denominator $\ell_d$ of ${\tilde L}^{nrc}$ in \eqref{equ:alpha_nonresonant} as a function of the sphere thickness $r_1$. When the metamaterial sphere is not resonant, the denominator $\ell_d$ can be either positive or negative: if $r_1<0.025$~m, $\ell_d<0$ while if $r_1>0.025$~m, $\ell_d>0$. Since $\ell_n$ does not change its sign, $L$ have different signs in the two regions.
As a result, the magnetic field generated by the coil should change its direction both inside and outside the shell. In Fig. \ref{fig:negative_inductance} and Fig. \ref{fig:positive_inductance}, we simulate the direction of magnetic field in Comsol. We observe that when $r_1=0.024$~m and $r_1=0.026$~m, the magnetic field have different directions, which validates that existence of negative self-inductance in M$^2$I.

As shown in Fig.~\ref{fig:3d_wi_t}, the real and negative self-inductance appears in a region on side of the resonance point (when $r_1<0.025$~m), where the second term (negative) in \eqref{equ:alpha_nonresonant} has a larger absolute value than $L_0$ (positive). When the sphere thickness $r_1$ becomes even smaller, the negative inductance is compensated by $L_0$ so that the total self-inductance becomes positive again. On the other side of the resonance point (when $r_1\geq0.025$~m), the self-inductance is always positive.

\begin{figure}[t]
\centering
%\vspace{20pt}
\subfigure[]{
    \includegraphics[width=1.64in]{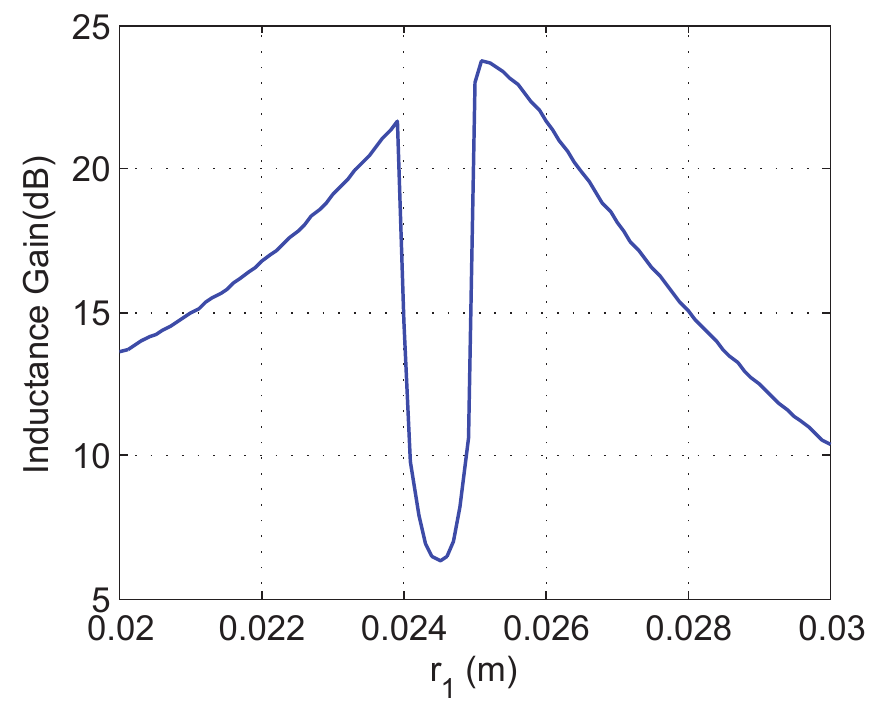}
     \label{fig:ig_without_match}}
\subfigure[]{
    \includegraphics[width=1.57in]{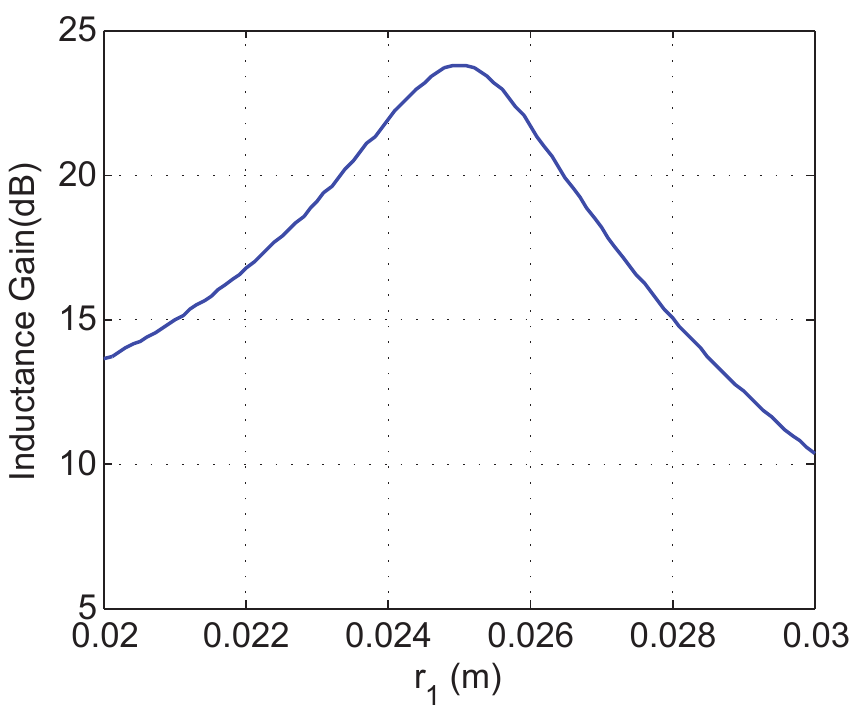}
  	\label{fig:ig_with_match}}
  \vspace{-5pt}
  \caption{The inductance gain $\mathcal{G}_M$ (a) without and (b) with the negative self-inductance matching. (distance is 5~m)}
  \vspace{-10pt}
  \label{fig:ig_with_without_match}
\end{figure}

Although the negative real self-inductance does not influence the metamaterial enhancement, it may incur significant loss in the MI coil circuit if not well designed. In M$^2$I transceiver, there are two types of resonance: the resonance in the metamaterial sphere and the resonance in the MI coil. The metamaterial sphere resonance is achieved by selecting optimal sphere thickness $r_1$ while the MI coil resonance is achieved by using compensation capacitor to cancel the impedance caused by the self-inductance. The negative real self-inductance cannot be compensated by capacitors, which incurs significant loss in the MI coil circuit. Fig. \ref{fig:ig_without_match} shows the inductance gain $\mathcal{G}_M$ as a function of the sphere thickness $r_1$ if the negative self-inductance is not compensated. We observe a significant performance deterioration on the one side of the resonance point (when $r_1<0.025$~m).

Since metamaterial is an effective medium, it's challenging to guarantee that the thickness $r_1$ exactly equals to the optimal value. If the fabricated $r_1$ is slightly smaller than the resonance point, significant performance drop can be incurred by the negative self-inductance. Two strategies can be adopted to address this problem.
First, the negative self-inductance can be canceled if we match it with a positive inductor. Fig. \ref{fig:ig_with_match} shows the inductance gain $\mathcal{G}_M$ as a function of the sphere thickness $r_1$ in the ideal case: if the self-inductance has a positive value, a capacitor is added to the coil circuit to compensate it; while if the inductance is negative, the capacitor is replaced with a positive inductor. We observe that the big drop in the inductance gain disappear. This solution requires the precise knowledge of the fabricated metamaterial sphere to determine whether to use compensation capacitor or compensation inductor.
%{\color{red} We can select whether to use compensation capacitor or compensation inductor once the metamaterial shell is fabricated.}%However,this solution requires a micro controller in the transceiver to adaptively select whether to use compensation capacitor or compensation inductor.
Second, a much simpler way to address the negative self-inductance problem is to fabricate the metamaterial sphere a little bit thicker than the optimal resonance point. As shown in Fig. \ref{fig:ig_without_match}, no drop of gain appears in the region that $r_1\geq 0.025$~m. Moreover, as discussed previously, the metamaterial enhancement in M$^2$I is not sensitive to the size deviation. Hence, a reliable M$^2$I system with good inductive gain can be derived if we design the sphere thickness $r_1$ slightly larger than the resonance value.

%In \eqref{equ:approx_m}, on the denominator there is a $r_1^2$ outside $[\det({\bf \widetilde{S}})]^2$. Hence, increasing $r_1$ can make the mutual inductance decreases faster than that by decreasing $r_1$. Accordingly, the gain is not symmetric with respect to $r_1=0.025$m in Fig. \ref{fig:ig_with_match}.

%Up to this point, in the analytical model, we consider there is no metamaterial loss and the two coils are coaxially aligned. If the metamaterial loss is taken into account, an imaginary part is added to $\mu_2$. As a result, by using \eqref{equ:resonnace}, we can no longer let $det({\bf \widetilde{S}}_{meta})=0$ since this is the resonance condition for no loss case. Thus, $det({\bf{S}}_{meta})$ would be larger than before resulting in smaller $\alpha_i$ and $\beta_i$. In next section, we numerically show the effect of loss.

%So far, from the above in-depth analysis, we find that
%\begin{itemize}
%\item  When $r_1$ is smaller than the resonant radius, negative inductance can be obtained.
%\item  The resistance caused by the imaginary self-inductance is greatly amplified when the shell is resonant.
%\item  The mutual inductance gain is relatively robust as the $r_1$ changes and we can obtain the highest inductance gain when the shell is resonant.
%\end{itemize}

\section{Channel Characteristics of M$^2$I Communications}

Based on the analytical model derived in Section III, we investigate the channel characteristics of the M$^2$I communication through both numerical analysis and FEM simulation in this section. The path loss, communication range, bandwidth, and channel capacity of both the M$^2$I point-to-point and the M$^2$I waveguide communications are quantitatively analyzed in various environments. If not specially specified, the default system and environment parameters used in this section are the same as Section III-B.3.

%In addition to analyze the performance, this section also presents a summary of key points that a designer might use to optimize the performance of metamaterial enhanced MI communication.

\subsection{Point-to-point M$^2$I Communication}

\begin{figure*}[t]
  \centering
  \subfigure[$r_1=0.025$~m. ]{
    \label{fig:sim_dipole}
    \includegraphics[width=0.3\textwidth]{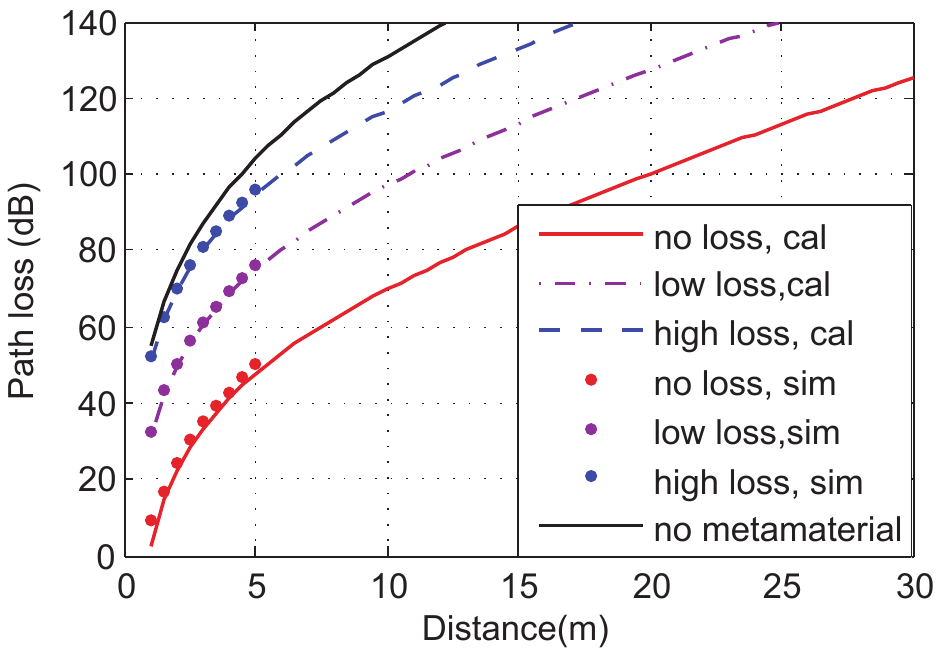}}\quad
  \subfigure[$r_1=0.027$~m.]{%
    \includegraphics[width=0.3\textwidth]{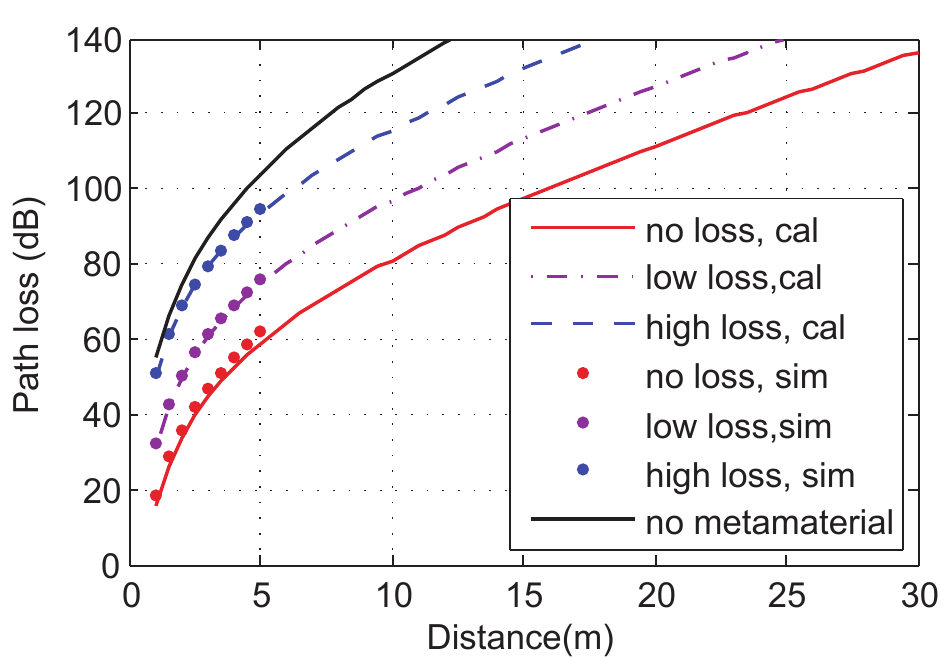}\quad
    \label{fig:sim_d_c}}
  \subfigure[$r_1=0.03$~m.]{%
    \includegraphics[width=0.29\textwidth]{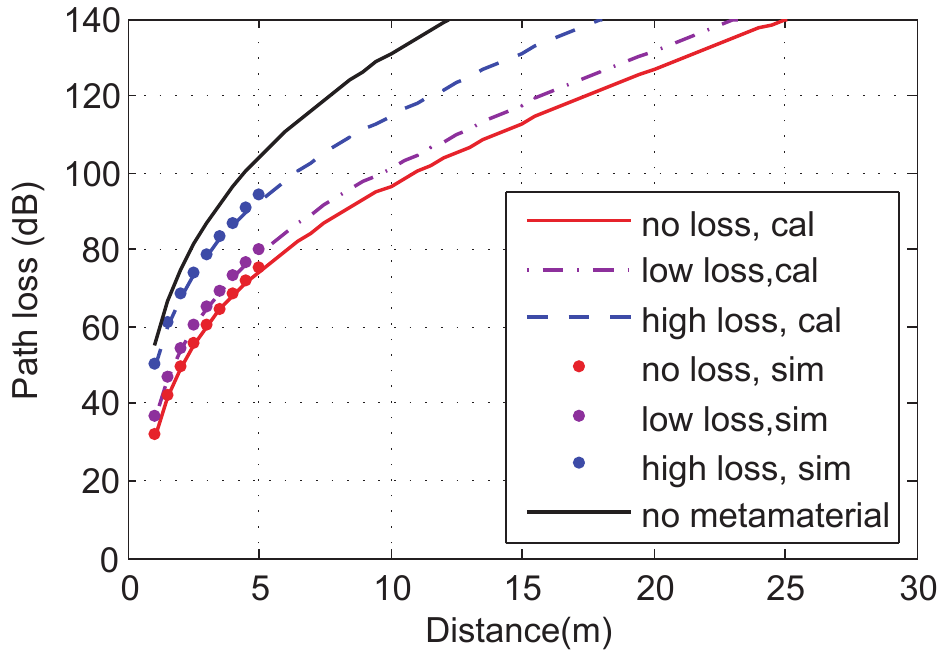}
    \label{fig:high_loss_pl}}
      \vspace{-5pt}
  \caption{Path loss of Point-to-Point M$^2$I communication.}
    \vspace{-5pt}
  \label{fig:sim_up1}
\end{figure*}
\begin{figure*}[t]
  \centering
  \subfigure[$r_1=0.025$~m. ]{
    \label{fig:p2p_bw_25}
    \includegraphics[width=0.3\textwidth]{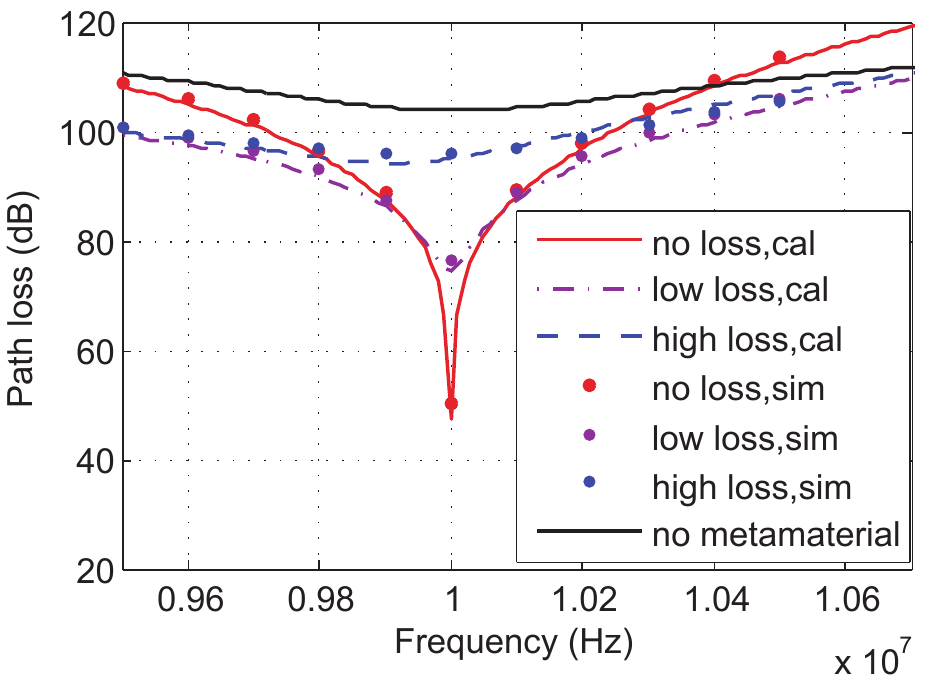}}\quad
  \subfigure[$r_1=0.027$~m.]{%
    \includegraphics[width=0.3\textwidth]{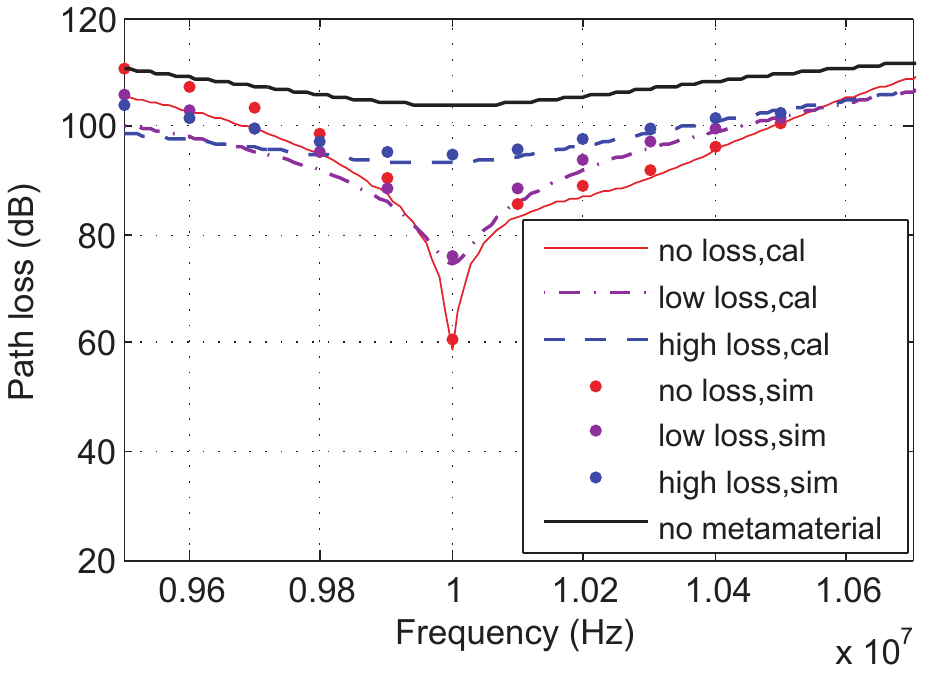}\quad
    \label{fig:p2p_bw_27}}
  \subfigure[$r_1=0.03$~m.]{%
    \includegraphics[width=0.29\textwidth]{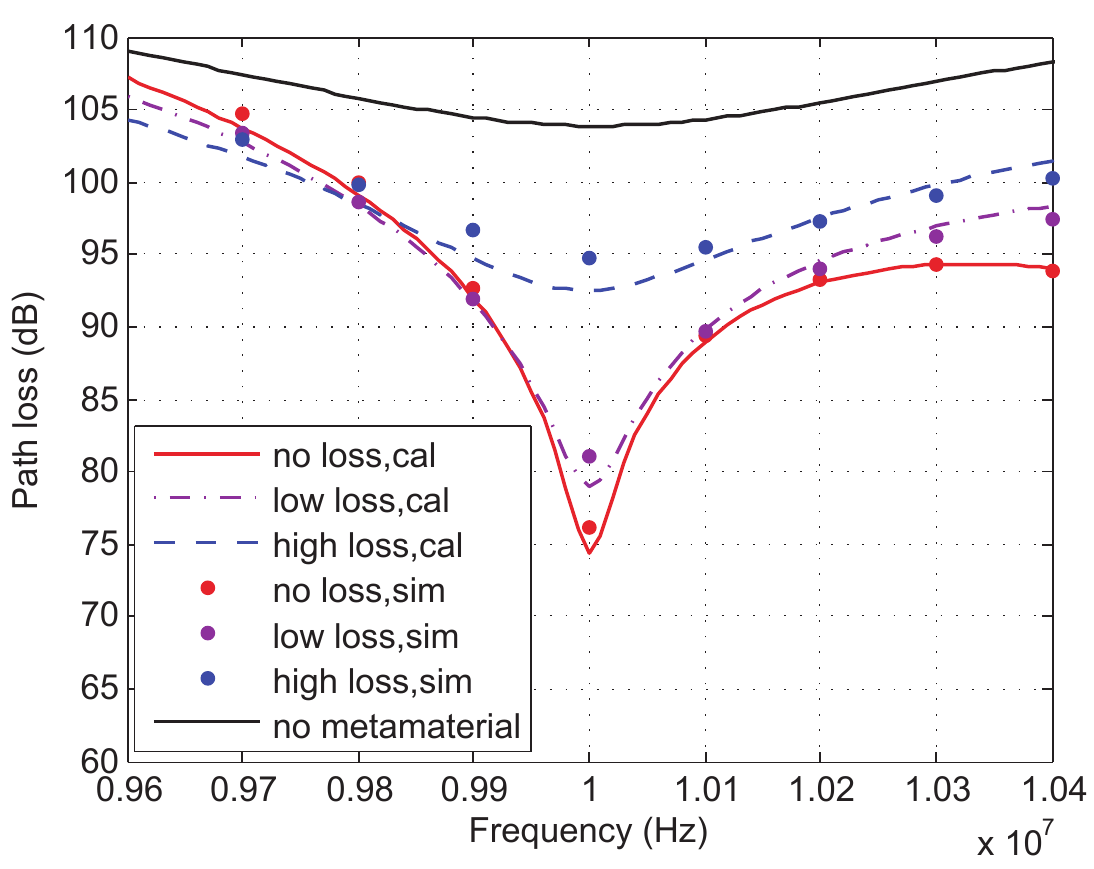}
    \label{fig:p2p_bw_30}}
      \vspace{-5pt}
  \caption{Frequency response of Point-to-Point M$^2$I communication (distance is 5~m).}
    \vspace{-5pt}
  \label{fig:p2p_bw}
\end{figure*}
\begin{figure*}[t]
  \centering
  \subfigure[$r_1=0.025$~m. ]{
    \label{fig:p2p_capacity_25}
    \includegraphics[width=0.3\textwidth]{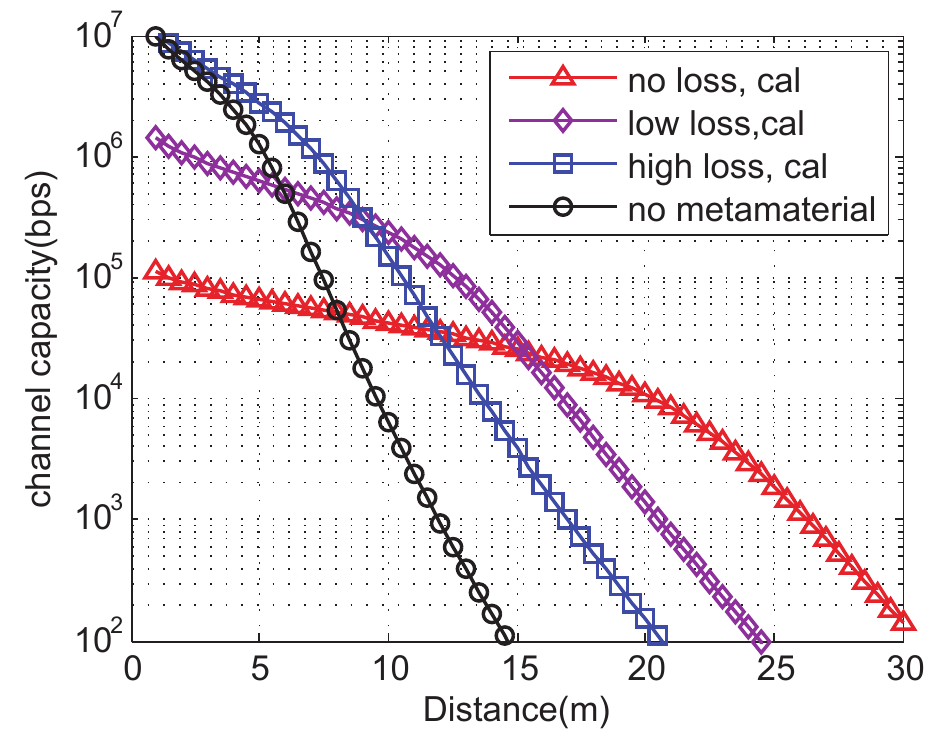}}\quad
  \subfigure[$r_1=0.027$~m.]{%
    \includegraphics[width=0.3\textwidth]{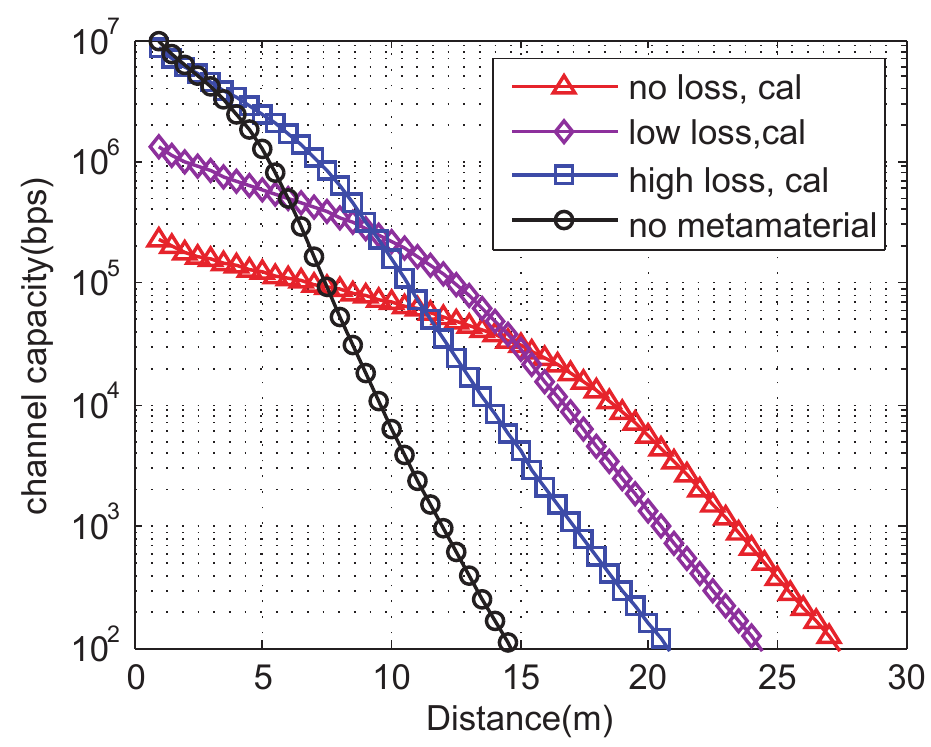}\quad
    \label{fig:p2p_capacity_27}}
  \subfigure[$r_1=0.03$~m.]{%
    \includegraphics[width=0.29\textwidth]{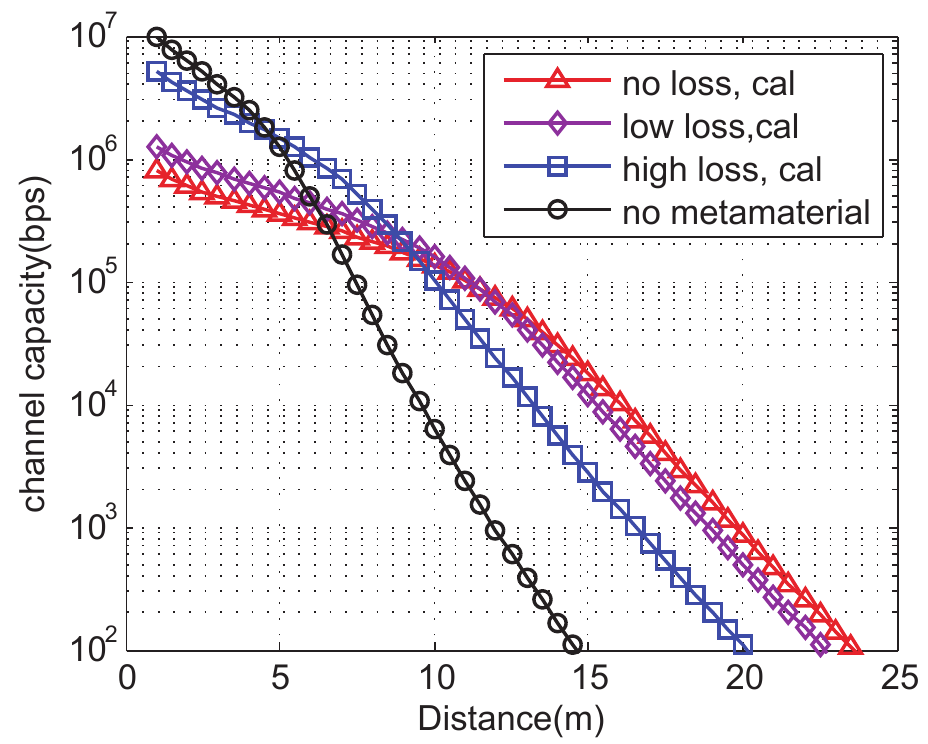}
    \label{fig:p2p_capacity_30}}
      \vspace{-5pt}
  \caption{Channel capacity of Point-to-Point M$^2$I communication.}
    \vspace{-5pt}
  \label{fig:p2p_channel_capacity}
\end{figure*}

\begin{figure*}[t]
  \centering
  \subfigure[Path loss. ]{
    \label{fig:path_concrete}
    \includegraphics[width=0.3\textwidth]{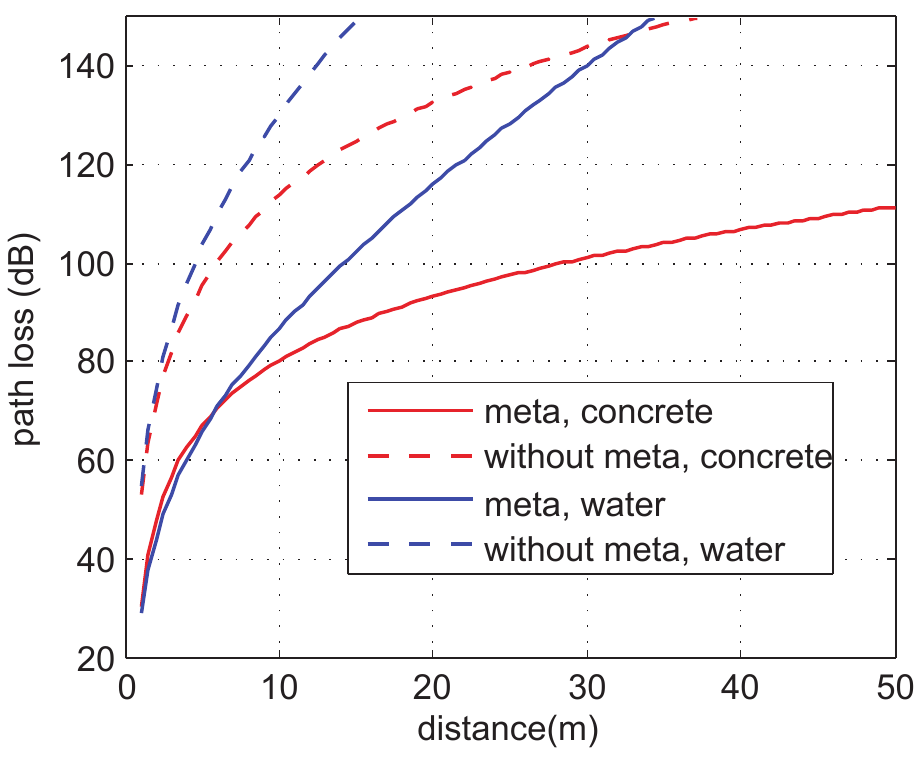}}\quad
  \subfigure[Frequency response (distance is 10 m).]{%
    \includegraphics[width=0.3\textwidth]{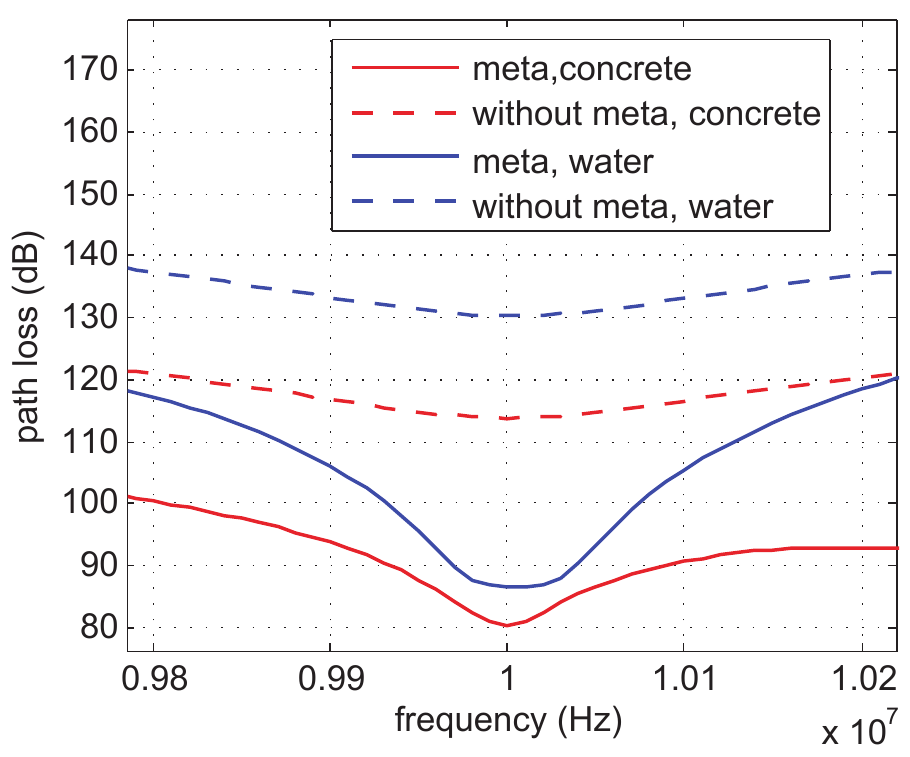}\quad
    \label{fig:bw_concrete}}
  \subfigure[Channel capacity]{%
    \includegraphics[width=0.29\textwidth]{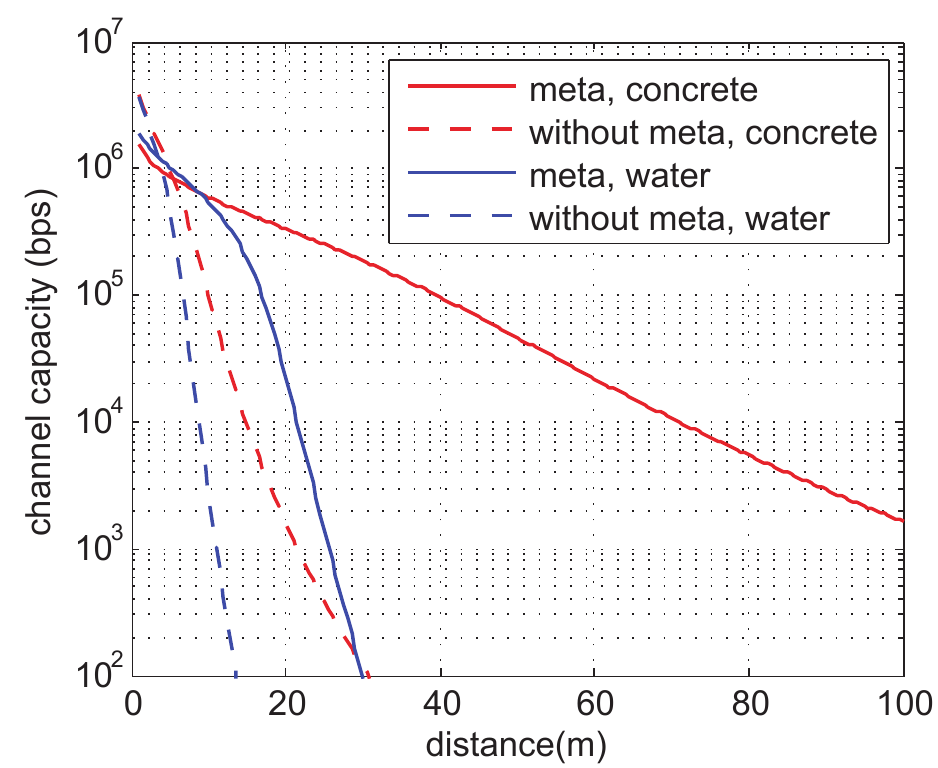}
    \label{fig:capacity_concrete}}
      \vspace{-5pt}
  \caption{Channel characteristics of M$^2$I in concrete ($\sigma$=0.1~mS/m) and water ($\sigma$=10~mS/m). Metamaterial has low loss ($\mu_2$=$(-1-0.005j)\mu_0$).}
    %\vspace{-5pt}
  \label{fig:other_medium}
\end{figure*}

\begin{figure*}[t]
  \centering
  \subfigure[Path loss. ]{
    \label{fig:wg_pathloss}
    \includegraphics[width=0.3\textwidth]{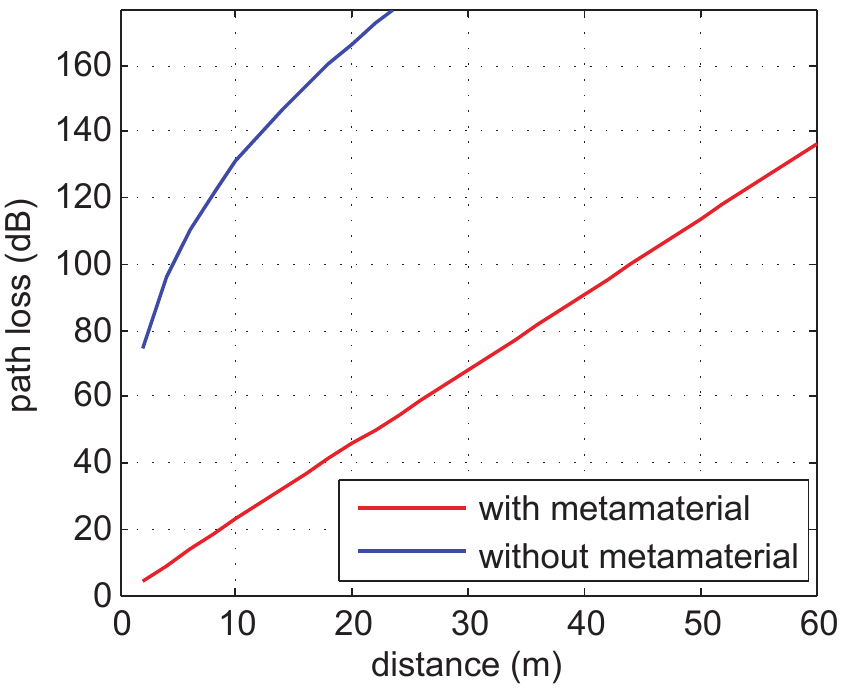}}\quad
  \subfigure[Frequency response (distance is 20 m).]{%
    \includegraphics[width=0.3\textwidth]{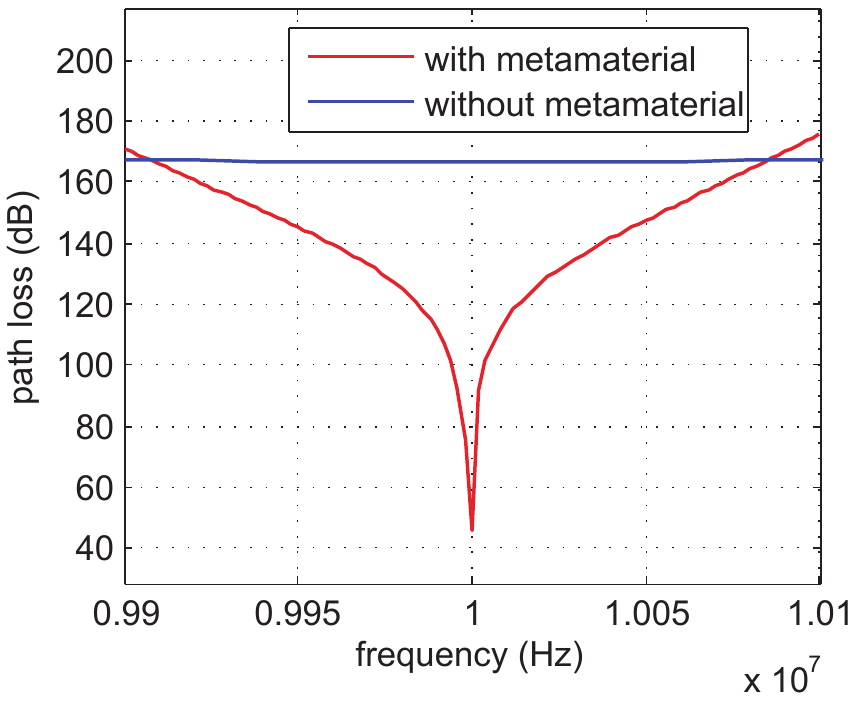}\quad
    \label{fig:wg_bw}}
  \subfigure[Channel capacity.]{%
    \includegraphics[width=0.29\textwidth]{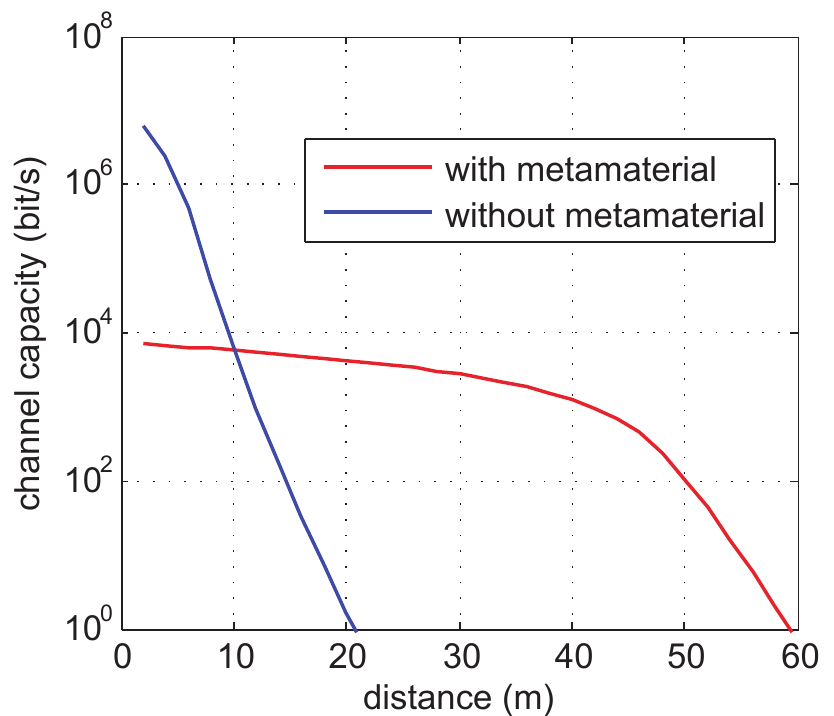}
    \label{fig:wg_capacity}}
      \vspace{-5pt}
  \caption{Channel Characteristics of M$^2$I waveguide.}
    \vspace{-10pt}
  \label{fig:sim_up}
\end{figure*}

%\subsubsection{Path loss}
\begin{figure}[t]
  \centering
    \includegraphics[width=0.46\textwidth]{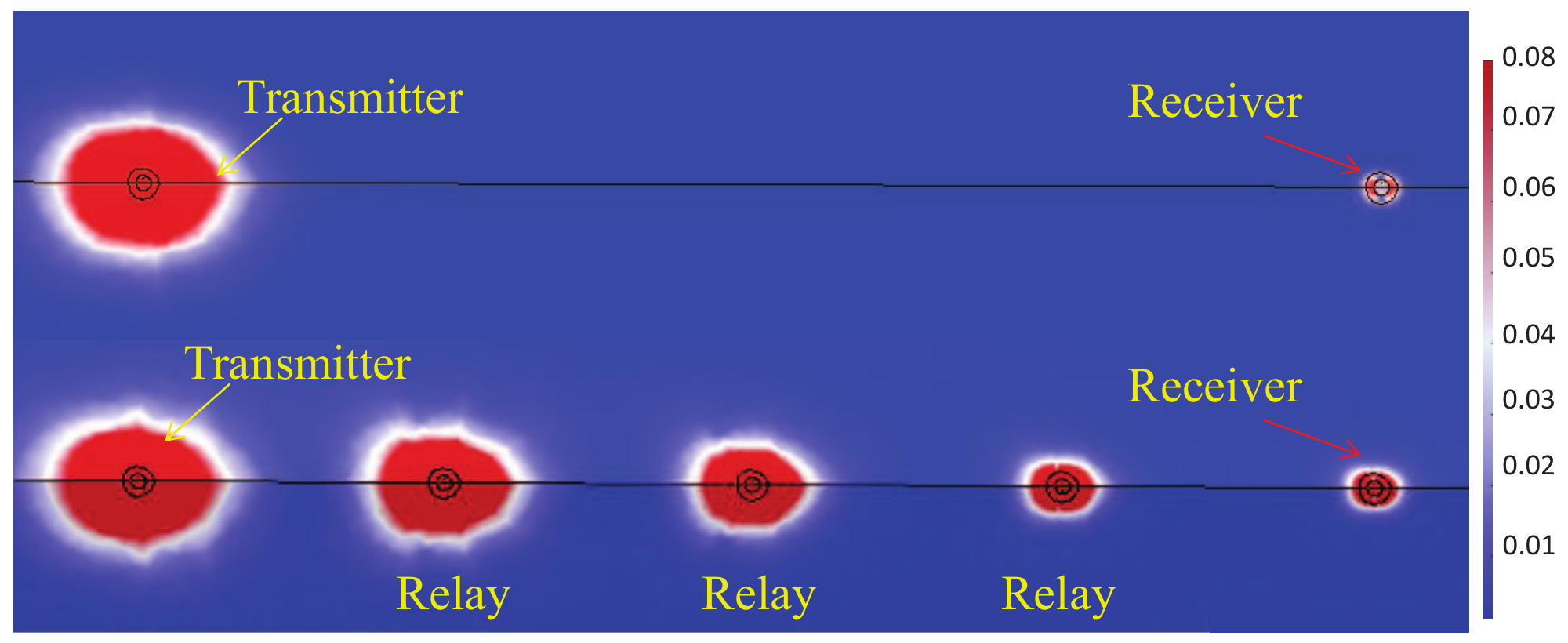}
    \vspace{-5pt}
  \caption{Magnetic field (Unit: A/m) of metamaterial-enhanced point-to-point M$^2$I communication (upper) and M$^2$I waveguide (lower). The transmitter is the left most node and the receiver is the right most node. The distance between transmitter and receiver is 4~m. In the waveguide, the interval of each two adjacent nodes is 1~m.}.
  \vspace{-0pt}
  \label{fig:wg_illustration}
\end{figure}
Fig. \ref{fig:sim_up1} shows the point-to-point M$^2$I path loss as a function of the communication distance through both theoretical calculation and FEM simulation. Similar to Section III-B.3, three levels of metamaterial loss are compared, including the no loss ($\mu_2=-\mu_0$), the low loss ($\mu_2=-\mu_0-j0.005\mu_0$), and the high loss ($\mu_2=-\mu_0-j0.05\mu_0$). To test the system robustness to the practical fabrication, three metamaterial sphere thicknesses are compared, including the resonance size ($r_1=0.025$~m) as well as two larger sizes ($r_1=0.027$~m and $r_1=0.03$~m). The sphere that is thinner than the resonance size is not considered due to the negative self-inductance problem discussed in Section III-C.2. The FEM simulation only shows the path loss within the distance of 5~m due to the high computation complexity of the high resolution simulation.
%{\color{red} the wire radius of the coil can be smaller than the mesh size which resulting in inaccurate results if we further increase the distance.}%longer simulation region requires very long computation time.
Consistent with the analysis in Fig. \ref{fig:ig_with_match}, resonant metamaterial sphere achieves the lowest path loss. As the thickness $r_1$ deviates from resonant radius, the gain introduced by metamaterial sphere gradually decreases. However, even $r_1$ is increased by 5~mm, the path loss of M$^2$I is still 30~dB lower than without the shell at 10~m distance when there is no loss in metamaterial which is shown in Fig. \ref{fig:high_loss_pl}.
Moreover, Fig. \ref{fig:sim_up1} shows that higher metamaterial loss can dramatically increases the M$^2$I path loss, which is consistent with the field analysis in Fig.~\ref{fig:txfield} and Fig.~\ref{fig:rxfield}. However, the influence of the metamaterial loss become less significant when the sphere thickness deviates from the resonant size, since both the thickness deviation and the metamaterial loss prevent the resonance in metamaterial sphere.
%The reason is that with loss the shell is hard to achieve resonance since $\det({\bf S}_{meta})$ cannot be minimized by \eqref{equ:resonnace}. Thus, when $r_1=0.025$m the determinant is much larger than that without loss. When the metamaterial has high loss, almost none of the the three configurations can significantly reduce $det({\bf {S}}_{meta})$. Hence, the gain is not as large as that without loss.
It should be noted that there are many ways to reduce the metamaterial loss, such as geometric tailoring \cite{Durdu_meta_loss} and high inductance-to-capacitance ratio \cite{zhou2008efficient}. In addition, by using the active metamaterials, such loss becomes even controllable \cite{Kurter_active}.

%On the one hand, the loss may reduce the performance of $M^2I$ communication. On the other hand, it can make the system more robust. In other words, even there are some errors during fabrication process, we can still obtain similar gain. For instance, consider that it's hard to make $r_1$ exactly 0.025m, but if $r_1=0.03$m, the system have the similar performance. In general, there is a tradeoff between robustness and high efficiency.

%\subsubsection{Bandwidth}
Fig. \ref{fig:p2p_bw} shows the theoretical and simulated frequency response of the same M$^2$I point-to-point system, where the 3 dB bandwidth can be read from the curves.
Since metamaterials are dispersive, we can only realize $\mu_2=-\mu_0$ at a narrow band. In order to conduct a more practical analysis, here we consider the Drude model \cite{Engheta_Meta} to model such dispersion, where the permeability $\mu_2$ is a function of frequency:
\begin{align}
\label{equ:mu2}
\mu_2(\omega)=\mu_0\left(1-\frac{\omega_{pm}^2}{\omega(\omega-j\Gamma_m)}\right),
\end{align}
where $\omega_{pm}$ and $\Gamma_{m}$ are the plasma and damping frequency, respectively. In this paper,  $\omega_{pm}$ is set as $8.89\times 10^7 $~rad/s, and $\Gamma_{m}$ is set as 0, $1.57\times10^5$~rad/s and $1.57\times10^6$~rad/s for no loss, low loss, and high loss at 10~MHz, respectively. The $\mu_2$ derived from \eqref{equ:mu2} is used in both the theoretical model and Comsol Multiphysics and the results are shown in Fig. \ref{fig:p2p_bw}. We observe that the bandwidth in M$^2$I communication is much narrow than the original MI system, which is due to the strong resonance introduced by metamaterials, especially in the no loss case.
%In addition, we can learn that the metamatreial loss has an adverse affect on the resonance. Consider Fig. \ref{fig:p2p_bw_25}, the loss increases the path loss meanwhile it makes the bandwidth broader.
As the metamaterial loss increases or the sphere thickness deviates from the resonance size, the system bandwidth increases. Hence, there exists a tradeoff between the low path loss and high bandwidth in M$^2$I.
%Moreover, the deviation from the resonant radius has the same affect on the bandwidth.

%\subsubsection{Channel Capacity}
%From previous discussion, we can see in $M^2I$ communication the lower path loss cannot guarantee a broader bandwidth. Usually, a strong resonance would generate a low path loss and narrow bandwidth, vice versa.

Since the objective of the M$^2$I communication system is to achieve a high data rate within a long transmission distance, the Shannon Capacity \cite{proakis2007digital,proakis2002communication} is used as the metric to evaluate the overall performance of the M$^2$I system:
%\begin{align}
%\mathcal{C}={\color{red}\int_{f_0-B/2}^{f_0+B/2} log_2\left(1+\frac{P_t \hat{\mathcal{L}}_{p2p}(f)}{N_{noise}}\right)\cdot df},
%\end{align}
\begin{align}
\mathcal{C}=\int_{f_0-B/2}^{f_0+B/2} log_2\left\{1+SNR(f)\right\}\,df,
\end{align}
where $f_0$ is the resonant frequency; $B$ is the 3 dB bandwidth; $SNR(f)=\frac{P_t \hat{\mathcal{L}}_{p2p}(f)}{N_{n}}$ is signal to noise ratio, where $P_t$ is the transmission power density, $\hat{\mathcal{L}}_{p2p}(f)$ is the antilogarithm of \eqref{equ:p2p_approx}, and $N_{n}$ is the noise power density. Since the bandwidth in M$^2$I is very small (in the order of KHz), the noise power density can be considered as a constant. Similarly, the density of $P_t$ is also a constant within the bandwidth. Fig. \ref{fig:p2p_channel_capacity} shows the channel capacity of the point-to-point M$^2$I system with different metamaterial loss and sphere thickness. We set the transmission power $P_t \cdot B$ as 10~dBm. It has been reported in \cite{Uribe_2009} that the power of underwater magnetic noise at MHz band is around -140~dBm. However, considering the underwater environment has relatively low background noise level, we set the noise power $N_n \cdot B$ as -100~dBm in this paper to guarantee the performance in much worse scenarios.
We observe that the M$^2$I system can reach the communication range of almost 30~m with kbps level data rate, which doubles the range of the original MI system. Even with metamaterial loss, the range can still exceed 20~m. In the near region, the bandwidth imposes a strong constraint on the capacity since the path loss is low enough. As the distance becomes larger, path loss plays a more important role and the advantages of M$^2$I become obvious.

\subsection{M$^2$I in Other Complex Environments}
Even most of the natural materials have the same permeabilities, their permittivities and conductivities can be dramatically different. Hence, we evaluate the performance of M$^2$I in other complex environments in the envisioned applications, including concrete and water. Different from soil, concrete has lower conductivity, while water has much larger permittivity and conductivity. In the numerical results, we consider concrete's relative permeability, relative permittivity ,and conductivity as 1, 4.5, and 0.1~mS/m, respectively. Water's relative permeability, relative permittivity, and conductivity are 1, 80.1, and 10~mS/m, respectively. In addition, the metamaterial has low loss and the shell has resonant inner radius ($r_1=0.025$~m). The path loss, bandwidth, and channel capacity of M$^2$I in concrete and water are shown in Fig. \ref{fig:other_medium}.
We observe that M$^2$I performs much better than conventional MI in both concrete and water, in aspects of communication range and channel capacity. In particular, with 0.1~mS/m conductivity in concrete, M$^2$I can achieve 100~kbps data rate at 40~m, while the original MI can only transmit in the same data rate within 10~m. If the conductivity in the medium is even lower, the communication range and data rate of the M$^2$I system can be further increased.

\subsection{M$^2$I Waveguide}
%According to \eqref{equ:wg_pathloss} we can see, if we increase the number of relay coils i.e. $n$, $1-n$ is enlarged. It seems we increased the path loss. However, more relay coils results in strong adjacent coupling. Thus $M$ is increased which can reduce the path loss. Consider an ideal case where $\frac{\omega|M|}{R_c+\omega L_i}\simeq1$,  the path loss is almost 0, which means the power can be delivered without loss.
The M$^2$I waveguide can be formed when multiple M$^2$I devices are placed along a line and the inter-distance between adjacent devices is small enough. For example, in the application of wireless sensor networks, many M$^2$I sensor nodes can be densely deployed. Between the M$^2$I transmitter and receiver, multiple M$^2$I nodes exist and form a M$^2$I waveguide along the transmission path.
Based on the M$^2$I channel model derived in Section III, we evaluate the performance of M$^2$I waveguide in this subsection.
%Noting that when we place many metamaterial shells in a certain area, the multiple scattering should be considered to get the most accurate model. However, the scattered fields are very weak and decrease fast. Since we don't concentrate at the field within 1m to the shells, we can safely ignore them.

Fig. \ref{fig:wg_pathloss} shows the path loss, bandwidth, and channel capacity of the M$^2$I waveguide and the original MI waveguide.
The thickness of the metamaterial sphere $r_1$ is fixed at the resonance size $0.025$~m and the no loss case is considered. Other configurations are the same as the point-to-point M$^2$I. The interval between adjacent M$^2$I device is 1~m. According to Fig. \ref{fig:wg_pathloss}, the M$^2$I waveguide can further increase the communication range compared with the point-to-point M$^2$I. Compared with the original MI waveguide, M$^2$I waveguide has much lower path loss but also much narrower bandwidth due to the joint resonant effects of multiple M$^2$I devices. According to the channel capacity given in Fig.~\ref{fig:wg_capacity}, without metamaterial, the larger coil formed waveguide cannot reach a communication range larger than 15~m. In contrast, the M$^2$I waveguide achieves the communication range of more than 40~m with the data rate at kbps level.

Fig. \ref{fig:wg_illustration} shows the Comsol simulations of the magnetic fields of the point-to-point M$^2$I communication and the M$^2$I waveguide. It's clear that with the help of the three passive relays, the magnetic field at receiver of the M$^2$I waveguide is much larger than the point-to-point case. As a result, the signal power at the receiver in M$^2$I waveguide can be increased.

\section{Conclusion}
In this paper, the metamaterial-enhanced magnetic induction (M$^2$I) communication mechanism is proposed for wireless applications in complex environments. An analytical channel model is developed to lay the foundation of M$^2$I communications and networking under the impacts from lossy transmission medium. The channel model reveals unique properties of M$^2$I communications, including the negative self-inductance and frequency-dependent resistance, which provides principle and guidelines in the joint design of communication systems and metamaterial antennas. The proposed M$^2$I mechanism and the channel model are validated and evaluated by using both the FEM simulations and proof-of-concept experiments. The results of this paper confirm the feasibility of achieving tens of meters communication range in M$^2$I systems by using pocket-sized antennas.

%From our analysis, we find that the resonant shell can greatly amplify the magnetic field generated by the coil inside it. Meanwhile, the resistance of the coil is also increased by due to the reflected impedance from the lossy environment and metamaterial. In addition, with the help of metamaterial, we can achieve negative inductance with a certain configuration. More importantly, if the metamaterial has loss, the system is more robust. In other word, if the radiuses or permeability are not exactly follow the design, we can still obtain exciting gain.

%
%\section*{Acknowledgment}
%This work is based upon work supported by the US National Science Foundation (NSF) under Grant No. 1547908.
%

\section*{Appendix}

\subsection{Magnetic Field around Receiver}
The excitation source is the coil. Without metamaterial shell, the radiated fields can be expressed as \cite{Balanis_a},
\begin{align}
\label{equ:dipole_spherical}
\begin{cases}
{\bf h}_{r}=\frac{j k a^2 I_0 \cos{\theta}}{2r^2}\left[1+\frac{1}{jkr}\right]e^{-jkr}{\hat r};\\
{\bf h}_{\theta}=\frac{- k^2 a^2 I_0 \sin{\theta}}{4r}\left[1+\frac{1}{jkr}-\frac{1}{(kr)^2}\right]e^{-jkr}{\hat \theta};\\
{\bf e}_{\phi}=\eta \frac{k^2 a^2 I_0 \sin\theta}{4r}\left[1+\frac{1}{jkr}\right]e^{-jkr}{\hat \phi};\\
{\bf h}_{\phi}=0;{\bf e}_{r}=0;{\bf e}_{\theta}=0,
\end{cases}
\end{align}

The magnetic field inside and scattered by the shell can be expressed by \eqref{equ:layer_equation}. Also, the radiated magnetic field can be found in \eqref{equ:dipole_spherical}. By enforcing the boundary conditions and rearranging the items we can find
\begin{align}
{\bf \Psi}_t=
\begin{pmatrix}
-\frac{\omega^2\sqrt{ \mu_1^3 \epsilon_1} a^2 I_0}{4r_1}\left[1+\frac{1}{jk_1 r_1}\right]e^{-jk_1r_1}\\
\frac{j \omega^3 \mu_1^2 \epsilon_1 a^2 I_0}{4}\left[1+\frac{1}{jk_1r_1}-\frac{1}{(k_1r_1)^2}\right]e^{-j k_1 r_1}\\
0\\
0
\end{pmatrix}\;.
\end{align}
and ${\bf S}_{meta}$ (shown on the top of next page).
%\newcounter{mytempeqncnt}
\begin{figure*}[t]
\vspace{-5mm}
\normalsize
\begin{equation}
\footnotesize
\label{equ:coefficient_matrix}
{\bf S}_{meta}\!=\!
\begin{pmatrix}
j_1(k_1 r_1)\!&-j_1(k_2 r_1)\!& -y_1(k_2 r_1)\!&0\\
j_1(k_1 r_1)\!+\!k_1 r_1 {j_1}'(k_1 r_1)\!& -\frac{\mu_1}{\mu_2}[j_1(k_2 r_1)\!+\!k_2 r_1 {j_1}'(k_2 r_1)]\!& -\frac{\mu_1}{\mu_2}[y_1(k_2 r_1)\!+\!k_2 r_1 {y_1}'(k_2 r_1)]\!&0\\
0 \!& j_1(k_2 r_2) \!& y_1(k_2 r_2) \!&-h_1^{(2)}(k_3 r_2)\\
0 \!& j_1(k_2 r_2)\!+\!k_2 r_2 {j_1}'(k_2 r_2) \!&y_1(k_2 r_2)\!+\!k_2 r_2 {y_1}'(k_2 r_2) \!&-\frac{\mu_2}{\mu_3}[h_1^{(2)}(k_3 r_2)\!+\!k_3 r_2 {h_1^{(2)}}'(k_3 r_2)]
\end{pmatrix}\!\!
\end{equation}
\hrulefill
\end{figure*}

The difference between the transmit coil and receive coil is the excitation source. As shown in Fig. \ref{fig:sys}, the magnetic field generated by the transmit coil is scattered on the second sphere. According to Mie theory, multiple mode decomposition is required to find the exact solution. Since the size of the metamaterial shell is much smaller than the signal wavelength in the envisioned applications (MHz band signal with pocket-sized device), the Rayleigh approximation can be applied \cite{4907079}. When a spherical scatter is much smaller than the wavelength, the first order of the Mie solution can be a good approximation to calculated the magnetic field.

Therefore, the format of the EM field intensity inside the receiver is the same as that around the transmitter. However, the coefficients in the formulas are different and need to be determined by the new boundary conditions. Since the shell is much smaller than wavelength, all the incoming magnetic fields on the shell can be assumed to have the same magnitude $h$. $h$ can be obtained from field in the third layer in \eqref{equ:thirdlayer}, i.e., ${\bf h}_{r3}$ and ${\bf h}_{\theta3}$. As shown in Fig. \ref{fig:sys}, we build a new spherical coordination whose origin is the center of the receiver and the magnetic field is along $z$ axis. Then, the magnetic field is decomposed along ${\hat r}$ and ${\hat \theta}$ direction, so that ${h}_{r}=-{ h}\cos\theta_0$ and ${ h}_{\theta}={h}\sin\theta_0$, where $\theta_0$ is the angle between the incoming magnetic field and ${\hat r}$.

Then we can obtain \eqref{equ:matrix_receive}. ${\bf S}_{meta}$ is the same as \eqref{equ:coefficient_matrix} and
\begin{align}
{\bf \Psi}_r=
\begin{pmatrix}
0\\
0\\
\frac{\omega r_2 \mu_3 h}{2 j}\\
-j \omega r_2 \mu_2 h
\end{pmatrix}\;.
\end{align}
By solving \eqref{equ:matrix_receive}, we can obtain all the unknown coefficient $\beta_i$.
\subsection{Subwavelength Approximation}
For those special functions, if $x<<1$, $j_1(x)\simeq \frac{x}{3}$, ${j_1}'(x)\simeq \frac{1}{3}$, $y_1(x)\simeq-\frac{1}{x^2}$, ${y_1}'(x)\simeq \frac{2}{x^3}$, ${h_1^{(2)}}(x)\simeq \frac{x}{3}+\frac{j}{x^2}$, and ${h_1^{(2)}}'(x) \simeq\frac{1}{3} -\frac{2j}{x^3}$. In the above approximations, we only keep the dominant real part and dominant imaginary part in the functions.

In addition, we consider there is no loss in the first layer and the wavenumber is real. According to the effective parameter analysis of the metamaterials in \cite{Alu_MNG}, the wavenumber in the second layer ($k_2$) is pure imaginary since the metamaterial adopted in this paper only has negative permeability. Moreover, since the environment is lossy (complex permittivity), the wavenumber of the propagation medium ($k_3$) is a complex number. Thus, $k_1=\rho_1$, $k_2=j\rho_2$, $k_3=\rho_{3r}-j\rho_{3i}$, all $\{\rho_x\}$ are real positive numbers,

By using the above approximations, \eqref{equ:coefficient_matrix} can be simplified as
\begin{equation}
\label{equ:simple_s}
{\bf S}_{meta}\approx{\bf \widetilde{S}}_{meta}=
\begin{pmatrix}
\frac{\rho_1 r_1}{3}&-\frac{j \rho_2 r_1}{3}&\frac{-1}{\rho_2^2 r_1^2}&0\\
\frac{2 \rho_1 r_1}{3}&-\frac{j 2 \rho_2 r_1 \mu_1}{3\mu_2}&\frac{\mu_1}{r_1^2 \rho_2^2\mu_2}&0\\
0&\frac{j \rho_2 r_2 }{3}&\frac{1}{\rho_2^2 r_2^2}&\zeta_1\\
0&\frac{j 2 \rho_2 r_2}{3}&\frac{-1}{\rho_2^2 r_2^2}& \zeta_2
\end{pmatrix}\,,
\end{equation}
where
\begin{subequations}
\begin{align}
\zeta_1\!=\!\frac{2 \rho_{3r} \rho_{3i}}{r_2^2(\rho_{3r}^2\!+\!\rho_{3i}^2)^2}\!-\!\frac{\rho_{3r} r_2}{3}\!+\!j\left[\frac{\rho_{3i} r_2}{3}\!-\!\frac{\rho_{3r}^2\!-\!\rho_{3i}^2}{r_2^2(\rho_{3r}^2\!+\!\rho_{3i}^2)^2}\right],
\end{align}
\begin{align}
\zeta_2=&\frac{-2 r_2 \mu_2 \rho_{3r}}{3 \mu_3}-\frac{2 \mu_2 \rho_{3r} \rho_{3i}}{r_2^2 \mu_3(\rho_{3r}^2+\rho_{3i}^2)^2} \nonumber \\
       &+j\left[\frac{2 r_2 \mu_2 \rho_{3i}}{3 \mu_3}+\frac{\mu_2(\rho_{3r}^2-\rho_{3i}^2)}{r_2^2 \mu_3(\rho_{3r}^2+\rho_{3i}^2)^2}\right].
\end{align}
\end{subequations}
Moreover, ${\bf \Psi}_t $ can be simplified as
\begin{equation}
\label{equ:simple_t}
{\bf \Psi}_t \approx {\bf \widetilde {\Psi}}_t=
\begin{pmatrix}
-\frac{j \omega \mu_1 a^2 I_0}{4 r_1^2}\\
\frac{j \omega \mu_1 a^2 I_0}{4 r_1^2}\\
0\\
0
\end{pmatrix}\;.
\end{equation}

With the simplified ${\bf \widetilde{S}}_{meta}$, the target coefficients $\alpha_1$ and $\beta_1$ can be explicitly formulated by using \eqref{equ:simple_s} and \eqref{equ:simple_t}. Then by substituting the solutions of $\alpha_1$ and $\beta_1$ into \eqref{equ:self_induction} and \eqref{equ:mutual_induction}, we derive the explicit expressions of the self-inductance \eqref{equ:approx_L} and mutual inductance \eqref{equ:approx_m} in M$^2$I communications.

Based on \eqref{equ:simple_s}, $\det({\bf \widetilde{S}}_{meta})$ can be given in \eqref{equ:det_approx}. Although the wavenumber and the radius of different layer may have different values and signs, their absolute values are in the same order. In \eqref{equ:det_approx}, the first term on the right side can be asymptotically approximated by $o(\frac{1}{{\bar k}^2{\bar r}^2})$. The second term is caused by the high order approximations of Bessel functions which are much smaller than $o(\frac{1}{{\bar k}^2{\bar r}^2})$ when $\bar k$ and $\bar r$ are smaller than 1.

\ifCLASSOPTIONcaptionsoff
 \newpage
\fi

\bibliographystyle{./IEEEtran}
\bibliography{revised_MetaCom}

\end{document}